\documentclass[onecolumn,sort&compress,numbers]{els-mrw} 

\usepackage{amsmath,amssymb,amsfonts,amsthm,makeidx,graphicx}
\usepackage{txfonts}
\usepackage{helvet}

\begin{document}


\chapter{Chiral perturbation theory}
\label{chap:CHPT}

\author[1,2,3]{Ulf-G. Mei{\ss}ner}%

\address[1]{\orgname{Universit\"at Bonn}, \orgdiv{Helmholtz-Institut f\"ur Strahlen- und Kernphysik and Bethe Center for Theoretical Physics}, \\
\orgaddress{D-53115 Bonn, Germany}}
\address[2]{\orgname{Forschungszentrum J\"ulich}, \orgdiv{Institute for Advanced Simulation (IAS-4)}, \orgaddress{D-52425 J\"ulich, Germany}}
\address[3]{\orgname{Tbilisi State University}, 
  \orgaddress{0186 Tbilisi, Georgia}}

\articletag{Chapter Article tagline: update of previous edition, reprint.}

\maketitle

\begin{abstract}[Abstract]
	In the limit of vanishing up, down and strange quark masses, QCD exhibits a chiral symmetry. This
	symmetry is broken spontaneously to its vector subgroup, giving rise to Goldstone bosons. These
	acquire a small mass through the explicit chiral symmetry breaking for non-vanishing quark masses.
        The consequences of these broken symmetries can
	be investigated  in a suitably tailored effective field theory called chiral pertubation theory. It admits 
	a perturbative expansion in the external momenta and the Goldstone boson masses and can be systematically
	analyzed in terms of a loop expansion. The appearing  ultraviolet divergences in loop diagrams can be dealt with 
	order-by-order	through the  Goldstone  boson contact interactions. Matter fields like
	the lowest-lying baryons can also  included, leading to a rich and testable phenomenology of low-energy
	QCD.  
\end{abstract}

\begin{keywords}
 	Chiral symmetry\sep Effective field theory\sep Goldstone bosons  \sep Chiral anomalies \sep Matter fields
\end{keywords}

\begin{glossary}[Nomenclature]
	\begin{tabular}{@{}lp{34pc}@{}}
		CHPT& Chiral perturbation theory\\
		EFT & Effective field theory\\
                 GB  & Goldstone boson\\
		LEC & Low-energy constant\\
		QCD & Quantum chromodynamics\\
		SSB & Spontaneous symmetry breaking\\
	\end{tabular}
\end{glossary}

\section*{Objectives}
\begin{itemize}
	\item In Sect.~\ref{sec:symm} the chiral symmetry of the strong interactions in the sector of the light
	quarks up, down and strange and its various
	breakings are discussed. The spontaneous breakdown of the chiral symmetry down to its vectorial
	subgroup leads to the appearance of massless Goldstone bosons. These can be identified with the
        low-lying octet of pseudoscalar mesons once the explicit chiral symmetry breaking through the
        light quark masses is accounted for. In the low-energy regime,
	an effective field theory (EFT) can be formulated in terms of these Goldstone bosons coupled to 
	matter fields and external sources. The basic rules to construct such an EFT are also discussed.
	\item Sect.~\ref{sec:meson} discusses the construction of the Goldstone boson EFT, that is
	chiral perturbation theory, including the
	power counting that maps the expansion in small momenta and meson masses onto a loop expansion.
	The emergence of low-energy constants, that parameterize the physics of shorter distances than
	the one from the Goldstone bosons, and the related physics is sketched.  The role of unitarity in the
	EFT  is also discussed. A few assorted applications
	to show the power of this framework are given.
      \item Sect.~\ref{sec:baryon} deals with the inclusion of matter fields into the effective Lagrangian.
        In the two-flavor case, this is the nucleon doublet and for three flavors, one deals with
        the low-lying baryon octet. The basic steps to couple the matter fields in a chiral-invariant fashion to the
	Goldstone bosons are sketched. The appearance of a large scale, the baryon mass, upsets the power counting.  
	Three methods are discussed that allow for a treatment of the baryon mass such that the power
	counting is restored. Two interesting applications of this framework that have led to new insights into
	the chiral dynamics of the strong interactions are presented.
\end{itemize}

\section{Introduction}\label{intro}

Quantum Chromodynamics (QCD), the gauge theory of the strong interactions, exhibits some
remarkable features. Its basic constituents, the quarks and gluons, are not found in
isolation but rather in composite objects, the hadrons. This phenomenon is called color
confinement, and it has so far not  been understood. It renders first principles calculations
of QCD in the sector of the light quarks (up, down and strange) extremely difficult.
However, as will be shown in the following, we can formulate an effective field theory 
at low-energies based on the broken chiral symmetry of QCD that fulfills the same
Ward identities as the underlying gauge theory. This is called chiral perturbation theory (CHPT).
The chiral symmetry of the ground state, SU(3)$_L\times$SU(3)$_R$ is spontaneously broken
to its vectorial subgroup SU(3)$_V$, leading to eight massless excitations, the Goldstone
bosons. These can be identified with the pseudoscalar meson octet $(\pi^0,\pi^\pm, K^0, \bar{K}^0,
K^\pm, \eta)$ once one realizes that the small quark masses $m_u,m_d,m_s$ lead to small
Goldstone boson masses, where small refers to the typical hadronic scale of about 1~GeV.
Using the powerful tool of effective Lagrangians, the consequences of the spontaneous
and explicit chiral symmetry breaking can be analyzed to high precision in CHPT. In addition,
QCD exhibits anomalous chiral symmetry breaking that is generated by renormalization.
The consequences of this symmetry breaking can also be
investigated in this framework. Finally, matter fields like the nucleons (protons and neutrons
or, more generally, the low-lying baryon octet)  can also
be included leading to even more testable predictions of the chiral symmetry of QCD. In this
chapter, the basic ideas and principles underlying CHPT are discussed together with a few
representative applications. A much more detailed exposition of these topics can be
found in the textbook~\cite{Meissner:2022cbi}.

\section{Symmetries of light-quark QCD and their breakings}
\label{sec:symm}

QCD is a non-abelian gauge theory based on the local color SU(3)$_c$ symmetry.
Its Lagrangian is given by 

\begin{equation}
{\cal L}_{\rm QCD} = -\frac{1}{4} \, G_{\mu\nu}^a G^{\mu\nu, a} +
\sum\limits_{f} \bar q_f (i D\!\!\!/ - {\cal M} ) q_f + \ldots~, ~~~~a= 1,\ldots,8~,
\end{equation}
with $G_{\mu\nu}^a =\partial_\mu A_\nu^a - \partial_\nu A_\mu^a - g [A_\mu^b, A_\nu^c]$ the gluon field strength
tensor, $A_\mu^a$ collects the eight gluon fields and $D_\mu = \partial_\mu - ig A_\mu^a \lambda^a/2$ is the covariant
derivative, the $\lambda^a$ are the eight Gell-Mann matrices (the generators of the SU(3)$_c$ group)
and $g$ is the SU(3)$_c$ gauge coupling. The diagonal matrix ${\cal M}$ collects the quark masses
$m_u,m_d,m_s,m_c,m_b,m_t$ corresponding to the six observed  quark flavors. 
Not given are gauge-fixing terms required for a consistent renormalization of the theory and further, the
so-called $\theta$-term is omitted here. Upon renormalization, QCD generates
a scale through the running of its coupling constant $\alpha_s = g^2/4\pi$. This scale is called $\Lambda_{\rm QCD}
\simeq 240\,$MeV in the standard $\overline{\rm MS}$ renormalization scheme at the scale $\mu =2\,$GeV. The
quark masses are usually determined within the same scheme and at the same scale (for a renormalization-scale
independent definition, see e.g.~\cite{Gasser:1982ap}). It is found that the up, down
and strange quarks have masses well below $\Lambda_{\rm QCD}$, whereas the mass of the remaining quarks 
are much larger than $\Lambda_{\rm QCD}$. For their actual values, see~\cite{ParticleDataGroup:2024cfk}.  
Therefore, QCD separates into two sectors, called the {\bf light} 
and the {\bf heavy} quark sector, respectively. 
In what follows, we will only consider the light quark sector, with
the heavy quarks integrated out. The QCD Lagrangian can be written as
\begin{BoxTypeA}[sec2:box1]{}
\begin{equation}
{\cal L }_{\rm QCD} = {\cal L }_{\rm QCD}^0 - \bar
q {\cal M} q~,~~~~{q=\begin{pmatrix}u\\d\\s\end{pmatrix}}~,~~~~~
~{\cal M}=\begin{pmatrix} m_u & &\\  & m_d &\\ & & m_s\end{pmatrix}~.
\end{equation}
\end{BoxTypeA}
\noindent
 This theory has remarkable symmetry, namely in the {\bf chiral limit} of vanishing
quark masses, $m_u=m_d=m_s=0$,  ${\cal L}^0_{\rm QCD}$  exhibits a chiral symmetry. This means that the left- ($L$) 
and right-handed ($R$) quark
fields decouple and can be rotated independently from each other, 
\begin{eqnarray}
{\cal L }_{\rm QCD}^0 (G_{\mu\nu} , q', D_\mu q') &=&
{\cal L }_{\rm QCD}^0 (G_{\mu\nu} , q, D_\mu q)~,\\
q &=& \frac{1}{2}(1-\gamma_5) q +   \frac{1}{2}(1+\gamma_5)  q= P_L  q + P_R q  = q_L + q_R~,\\
q' &=& RP_R q + LP_Lq = Rq_R + Lq_L~, ~~~~R, L \in {\rm SU(3)}_{R,L}~,
\end{eqnarray}
with $P_{L,R}^2 = P_{L,R}^{}$ and $P_L \cdot P_R = 0$. This global  symmetry is given by the group product 
SU(3)$_L \times$SU(3)$_R$.  By Noether's theorem, this leads to 16 conserved left- and right-handed
(or 8 vector $V_\mu^a$ and 8 axial-vector $A_\mu^a$) currents, $J_{L,R}^{\mu,a} = 
\bar{q}_{L,R} \gamma^\mu (\lambda^a/2)q_{L,R}$, 
$a = 1, \ldots, 8$, with $\partial_\mu J_{L,R}^{\mu,a} = 0$ (or, similarly, for $V^\mu = J_L^\mu +
J_R^\mu, ~~A^\mu = J_L^\mu - J_R^\mu$). From conserved currents, we can construct 8 conserved 
vector and 8 conserved axial-vector charges, $Q^a$ and $Q_5^a$, respectively.

In addition, one also observes a U(1)$_V$ symmetry, which counts the number of
quarks minus antiquarks, thus corresponding to baryon number. Classically, there is also a U(1)$_A$ symmetry, 
that is, however, broken as discussed below. As mentioned above, renormalization generates the scale $\Lambda_{\rm QCD}$.
In fact, classical QCD with vanishing quark masses exhibits a {\bf dilatation symmetry}.
More precisely, classical massless QCD is invariant under scale transformations
$x \to \lambda x, \psi(x) \to \lambda^{3/2} \psi(\lambda x),
A_\mu(x) \to \lambda A_\mu(\lambda x)$, with $\lambda \in \mathbb{R}\setminus\{0\}$. Consequently, there are
no massive states in QCD in this limit and this symmetry is anomalously broken, the so-called trace anomaly~\cite{Collins:1976yq}.
This has profound implication for the nucleon mass, see e.g the detailed discussion in Ref.~\cite{Donoghue:1992dd},
and is a beautiful manifestation of Wheeler's ``mass without mass''~\cite{Misner:1957mt}. 

\subsection{Spontaneous symmetry breaking}\label{sec:symmspon}

If the chiral symmetry would be the symmetry of the ground state, our world would look rather different,
because each particle would come with a partner of equal mass but opposite parity.  This can be most easily
understood in the so-called Wigner-Weyl realization, in which $Q_5^a |0\rangle = Q^a |0\rangle = 0 ~ (a=
1, \ldots , 8)$. The appearance of parity doublets follows from the observation that all these charges
commute with the QCD Hamiltonian, $[Q^a, H_{\rm QCD}] = [Q^a_5, H_{\rm QCD}] = 0$. 
Let us consider a single particle state  $ H_{\rm QCD} |\psi_p\rangle 
= E_p |\psi_p\rangle$ and perform an axial rotation, $ H_{\rm QCD}({\rm e}^{iQ_5^a}
|\psi_p\rangle) = {\rm e}^{iQ_5^a} H  |\psi_p\rangle = E_p ({\rm e}^{iQ_5^a} |\psi_p\rangle)$. We see that
the rotated state has the same energy $E_p$ as the original state but is of opposite parity due to the axial rotation.
This is obviously not the case in our world. In fact, the chiral symmetry is spontaneously broken, or hidden.  In terms of group theory, spontaneous symmetry breaking 
(SSB) is given by SU(3)$_L \times$SU(3)$_R \to$SU(3)$_V$, consistent with some general restrictions for
vector gauge symmetries (as it is the case for QCD)~\cite{Vafa:1983tf}). This SSB  has some remarkable consequences,
as it leads to the appearance of eight massless particles, the Goldstone bosons. The Goldstone theorem
\cite{Goldstone:1961eq} states that
for each broken generator, we have one massless particle. This can be understood in a nutshell (for a more rigorous
derivation, see e.g~\cite{Bernstein:1974rd}):  Let ${\cal H}$ be some Hamiltonian that is invariant
under some charges $Q^i$, i.e. $[{\cal H}, Q^i] = 0$, with $i = 1, \ldots, n$.
Assume further that $m$ of these charges ($m \leq n$) do not annihilate the
vacuum, that is $Q^j |0 \rangle \neq 0$ for $j =1, \ldots, m$.  Define a
single-particle state via $|\psi\rangle = Q^j |0 \rangle$.   This is an
energy eigenstate with eigenvalue zero, since $H |\psi\rangle  = H
Q^j|0\rangle=  Q^j H
|0\rangle = 0$. Thus, $|\psi\rangle$ is a single-particle state with $E =
\vec{p} = 0$, i.e. a massless excitation of the vacuum. These states are the
{\bf Goldstone bosons}, collectively denoted as pions $\pi(x)$ in what follows.
In QCD, we have  eight pseudoscaler massless Goldstone boson, as the vector SU(3) symmetry 
stays in tact. Through the corresponding symmetry current the Goldstone bosons couple 
directly to the vacuum,
\begin{equation}\label{GBME}
\langle 0 | J^0 (0) | \pi \rangle \neq 0~.
\end{equation}
In fact, the non-vanishing of this matrix element is a {\bf  necessary and
sufficient} condition for spontaneous symmetry breaking.  In QCD, this matrix element
is parameterized by the pion decay constant $F_\pi$, which thus constitutes an important
scale in QCD as it is an order parameter of SSB, more precisely
\begin{equation}\label{eq:axcurr}
\langle 0|A_\mu^i(0)|\pi^k(p)|0\rangle = i F_\pi p_\mu\delta^{jk}~,~~j,k=1,2,3,~~\mu = 0,1,2,3~,
\end{equation}
with $F_\pi \simeq 92\,$MeV. There are many other order parameters,
such as the vacuum expectation value (VEV) of the light-quark condensate 
\begin{equation}\label{eq:cond}
\langle  0 | \bar{q}q |0\rangle = \langle 0  | \bar q_L q_R + \bar q_R q_L|0\rangle~,
\end{equation}
which connects the left- and the right-handed quarks and is related to the number density of zero eigenvalues 
of the QCD Dirac operator~\cite{Banks:1979yr}. 
Another important property of Goldstone bosons is the derivative nature of their
coupling to themselves or matter fields. Again, in a hand-waving fashion, this
can be understood easily. As above, one can repeat the operation of acting
with the non-conserved charge  $Q^j$ on the vacuum state $k$ times, thus 
generating a state of $k$ Goldstone bosons that is degenerate with the
vacuum. Assume now that the interactions between the Goldstone bosons is
non-vanishing at zero momentum. Then, the ground state ceases to be degenerate
with the $k$ Goldstone boson state, thus the assumption must be incorrect.
Of course, this argument can also be made rigorous.
In the following, the derivative nature of the pion couplings will play an 
important role. Coming back to QCD, the eight lightest particles are indeed 
collected in the pseudoscalar octet, the three pions ($\pi^0, \pi^\pm$), the four kaons
($K^0, \bar{K}^0, K^\pm$) and the $\eta$ meson. So these would qualify as the Goldstone
bosons, if only they were massless. However, their masses are small compared to the
SSB scale $\Lambda_\chi \simeq 4\pi F_\pi \simeq 1\,$GeV~\cite{Georgi:1984zwz}.
One can also identify this scale with the lightest (narrow) non-Goldstone meson, the
$\rho(770)$ vector meson, that is $\Lambda_\chi \simeq 770\,$MeV.

The assumption of vanishing light quark masses in QCD defines the so-called {\bf chiral limit},
but there is one subtlety that needs to be mentioned. One needs to be careful to differentiate
between the three-flavor and the two-flavor chiral limit. In the latter, only the light up and down quarks
are massless, while the strange quark is considered heavy. This also has an influence on the so-called
low-energy constants (LECs) of the effective field theory (EFT), as discussed below. We end this
section with a quote: 
\begin{quote}
	\quotehead{QCD in the chiral limit}
	In this limit, QCD  is a theoreticians paradise: A theory without adjustable parameters.
	\source{Heiri Leutwyler~\cite{Leutwyler:1996et}}
\end{quote}

\subsection{Explicit symmetry breaking}\label{sec:symmexp}

As we have just seen,  the SSB of the chiral symmetry of QCD should lead to eight
massless pseudoscalar Goldstone bosons, that could be identified with the low-lying 
pseudoscalar meson octet. But the mesons in this octet are massive, so how can this be?
In fact, the light quarks are not massless but acquire small masses through the Higgs mechanism.
Clearly, such masses break the chiral symmetry {\bf explicitly}. However, QCD
possesses what is called an {\bf approximate} chiral symmetry. In that case, the mass spectrum
of the unperturbed Hamiltonian and the one including the quark masses can not be
significantly different. Stated differently, the effects of the explicit symmetry
breaking can be analyzed in perturbation theory.  This perturbation generates
the remarkable mass gap of the theory - the pions (and, to a lesser extent,
the kaons and the eta) are much lighter than all other hadrons. As we will show later,
to leading order in this explicit chiral symmetry breaking  the pion mass is given by
\begin{equation}\label{eq:GMOR}
M_\pi^2 = B\, (m_u+m_d)~,
\end{equation}
where the low-energy constant (LEC) $B$ is related to the quark condensate given in Eq.~\eqref{eq:cond},
and this equation also runs under the name of Gell-Mann--Oakes--Renner (GMOR) relation~\cite{Gell-Mann:1968hlm}.

The  quark mass term encodes further interesting symmetry. Many 
hadrons made of light quarks appear in isospin multiplets, characterized by very tiny
splittings of the order of a few MeV.  The most famous one is the isospin doublet made
of the proton and the neutron, with a mass difference $m_n-m_p = 1.3\,$MeV.
These splittings are generated by the small
quark mass difference $m_u -m_d$  and also by electromagnetic effects of similar
size (with the notable exception of the charged to neutral pion mass
difference that is almost entirely of electromagnetic origin). This is related to the
two-flavor quark mass term, namely for $m_u = m_d$, QCD is  invariant under SU(2) isospin 
transformations: 
\begin{equation}
q \to q'  = U q~,~~ q = \left(\begin{array}{cc} u \\ d\end{array}\right)~,~~~
U  = \left(\begin{array}{cc} a^* &  b^* \\ -b & a\end{array}\right)~,
~~~|a|^2+|b|^2 = 1~.
\end{equation}
In this limit, up and down quarks can not be disentangled as far as the
strong interactions are concerned.  The QCD quark mass term for the up and down
quarks can be decomposed into an isospin-conserving and an isospin-violating part,
\begin{equation}
{\cal H}_{\rm QCD}^{\rm SB} = m_u \,\bar u u + m_d
  \,\bar  d d = \dfrac{m_u+m_d}{2}(\bar u u + \bar d d)
              + \dfrac{m_u-m_d}{2}(\bar u u - \bar d d)~,
\end{equation}
where the first (second) term is an isoscalar (isovector).  Although the ratio
$(m_d-m_u)/(m_d+m_u)\simeq 1/3$ is not small, isospin is such a good 
symmetry because the true expansion parameter is $(m_d-m_u)/\Lambda_{\rm QCD}
\simeq 1/100$.

Extending these considerations to SU(3), one arrives at the eightfold way of
Gell-Mann and Ne'eman \cite{Gell-Mann:1964ewy,Neeman:1964rxv}
that played a decisive role in our understanding
of the quark structure of the hadrons. The SU(3) flavor symmetry
is also an approximate one, but the breaking is much stronger than it is
the case for isospin. From this, one can directly infer that the quark mass
difference $m_s - m_d$ must be much bigger than $m_d -m_u$.  We can also
infer this from the masses of the Goldstone bosons in QCD as shown below. Note that one 
often speaks of pseudo-Goldstone bosons for particles with a non-vanishing mass, 
but I will not use this jargon here.

\subsection{Anomalous symmetry breaking}\label{sec:symmanom}

There is one additional source of symmetry breaking in QCD, 
related to the fact that renormalization can break a symmetry
of the classical theory. This is most easily understood
in terms of the path integral representation of QCD. The effective
action contains an integral over the quark fields that can be expressed
in terms of the so-called fermion determinant. Invariance of the theory
under chiral transformations not only requires the action to be left
invariant, but also the fermion measure \cite{Fujikawa:1983bg}. 
Again, I only use a very hand-waving argument here and refer
to the quoted literature for the details.  A transformation of the fermion
field   can symbolically be written as $q \mapsto q' = {\rm e}^{-iS}q$, so that 
${\cal L}(q',\bar{q}',\ldots) = {\cal L}(q,\bar{q},\ldots)$, which means for the
path integral
\begin{equation}
\int [d\bar q][dq] \ldots \to |{\mathcal J}|\int [d\bar q'][dq'] \ldots~,
\end{equation}  
where $|{\mathcal J}|$ is the Jacobian of the transformation under consideration.
If the Jacobian is not equal to one, $|{\mathcal J}| \neq 1$,
one encounters a so-called  {\bf anomaly}.   Of course, such a statement has to be 
made more precise since the path  integral requires regularization and 
renormalization, still it captures the essence of the chiral anomalies
of QCD. One can show in general that certain 3-, 4-, and 5-point
functions with an odd number of external axial-vector sources are 
anomalous. As particular example, consider the divergence of the
singlet axial current,
\begin{equation}
\partial_\mu (\bar q \gamma^\mu \gamma_5 q) = 
\frac{N_f}{8\pi} G_{\mu\nu}^a \tilde{G}^{\mu\nu, a}~,
\end{equation}
where for the moment we consider $N_f$ massless quarks, and 
$\tilde{G}^{\mu\nu, a}$ is the dual gluon field strength tensor. Rescaling the gluon
field strength tensor as $G_{\mu\nu}\to g G_{\mu\nu}$, in order to get a standard normalization 
of the gluon field, one sees that the anomaly scales $1/N_c$, with $N_c$ the number
of colors~\cite{tHooft:1973alw}. In that limit, we thus could have 9 Goldstone bosons, but for finite $N_c$, the
anomaly generates  the $\eta '$  mass, i.e. this
particle stays massive in the chiral limit and thus does not qualify
as a ninth Goldstone boson~\cite{Weinberg:1975ui,Witten:1980sp,DiVecchia:1980yfw}.
There are many interesting aspects of anomalies in the context of QCD and chiral
perturbation theory. Here, I will only discuss a few anomalous processes and their
implications and refer to the review~\cite{Bijnens:1993xi}
that covers more of these issues.

\subsection{Effective field theory}\label{sec:EFT}

Effective field theory is a standard tool in many branches of physics, here we briefly discuss
its basic principles and how they are realized in the light-quark sector of QCD. Let me state this
with a quote from Weinberg's groundbreaking paper on effective Lagrangians:
\begin{quote}
	\quotehead{EFT theorem}
	Quantum field theory itself has no content beyond analyticity, unitarity, cluster decomposition, and 
	symmetry. This can be put more precisely in the context of perturbation theory: if one writes down the 
	most general possible Lagrangian, including all terms consistent with assumed symmetry
        principles, and then calculates matrix elements with this Lagrangian to any
        given order of perturbation theory, the result will simply be the most general
        possible S-matrix consistent with analyticity, perturbative unitarity, cluster
        decomposition and the assumed symmetry principles.
	\source{Steven Weinberg~\cite{Weinberg:1978kz}}
\end{quote}

This contains the basic principles of constructing an EFT. First, we must have {\bf scale separation}. That is,
if we want to study a theory in some low-energy region well below some scale $\Lambda$, then the physics
at scales above $\Lambda$ can not be resolved and can be parameterized in terms of contact interactions
 of the light degrees of freedom. These come with LECs, which are usually determined by a fit to
some low-energy data. Scale separation also implies that renormalizability in the strict sense is not applicable,
instead, the EFT can be renormalized order by order.
Related to the scale separation are the {\bf relevant degrees of freedom}, which are closely
related to the singularity structure of the low-energy S-matrix. In the case of QCD at low energies, say below the
scale of SSB, these are the pions (and, eventually, matter fields) and not quarks and gluons, that make up the
hadrons. The quark-gluon substructure can simply not be resolved in this energy regime. As we already noted,
{\bf symmetries} and their realization play an important role in the construction of the corresponding EFT. 
In low-energy QCD, this is naturally chiral symmetry breaking in its various forms as well as the discrete symmetries 
of parity, charge conjugation and time reversal (if the $\theta$-term is ignored as done here). Fourth, we need
a {\bf power counting} to organize the infinite tower of interaction terms we can construct from the underlying fields 
incorporating the relevant symmetries. Symbolically, any matrix element can be perturbatively expanded in powers of 
energies/momenta $q$ over the breakdown scale $\Lambda$ as
\begin{BoxTypeA}[sec2:box2]{}
\begin{equation}
    {\cal M} = \sum_\nu \left(\frac{q}{\Lambda}\right)^\nu f_\nu(q/\mu, g_i)~,
\end{equation}
\end{BoxTypeA}
\noindent
with $\mu$ some renormalization scale related to the required renormalization of loop diagrams
and the $g_i$ are the LECs. The index $\nu$ is bounded from below, which
allows for a systematic and controlled expansion. The smallest value of $\nu$ defines the
leading order (LO)  of the perturbative expansion in powers of $q/\Lambda$ and so on. Furthermore, $f_\nu$ is a function of
$O(1)$, and this property is called ``naturalness''. It is now believed that all quantum field  theories are
indeed EFTs, see e.g. Chap.~12 in \cite{Weinberg:1995mt}. We end this section with the remark that indeed it
has been shown that the chiral effective Lagrangian to be discussed below leads to the same Ward identities as QCD
\cite{Leutwyler:1993iq}, which puts Weinberg's theorem in this context on firm grounds.

\section{CHPT for mesons}\label{sec:meson}

\subsection{Effective Lagrangian}\label{sec:mesonL}

Based on the  Weinberg theorem, the low-energy theory of QCD corresponds to the mapping
\begin{equation}
{\cal L}_{\rm QCD}(q,\bar q, A) \to {\cal L}_{\rm eff}(U, \partial_\mu U,  \ldots, s,p,v,a)~,
\end{equation}
where $U$ collects the Goldstone bosons, $s,p,v,a$ are external scalar, pseudeoscalar, vector
and axial-vector  fields (in what follows, we will only consider $s$) and we have suppressed all possible
indices.  Accordingly, we now want to construct an EFT of the GBs and their interactions.
This program has been carried out in the groundbreaking papers of Gasser and Leutwyler,
which contain many more details than can be given here and should be consulted in any
case~\cite{Gasser:1983yg,Gasser:1984gg}.

It is most convenient
to collect the GBs in a matrix-valued field $U(x) $, which realizes a non-linear transformation
of the meson fields. In the three-flavor case, this reads [note that the argument $x$ in $U(x)$ will only be displayed when necessary]:
\begin{eqnarray}\label{eq:U}
U(x) = \exp( i \Phi (x)/F_\pi)~,~~ 
\Phi = \sqrt{2}\begin{pmatrix}
\pi^0/\sqrt{2} + \eta/\sqrt{6} & \pi^+ & K^+\\
\pi^- & -\pi^0/\sqrt{2} + \eta/\sqrt{6}  & K^0\\
K^- & \bar{K^0} & -2\eta / \sqrt{6} \end{pmatrix}~,
\end{eqnarray}
with $U$ a unitary matrix, $U^\dagger U= UU^\dagger =1$, that transforms under chiral symmetry as
\begin{equation}
U \mapsto RUL^\dagger~,~~L,R \in {\rm SU}(3)_{L,R}~.
\end{equation}
This clearly corresponds to a non-linear realization of the GB fields (for the general theory of
such non-linear realizations, see~\cite{Coleman:1969sm,Callan:1969sn}).
In the two-flavor case, this matrix reduces to a $2\times 2$ form in terms of the three pion fields,
so that the effects of the $\eta$ and the kaons are absorbed in the LECs. A first building block
that emerges naturally and embodies the derivative nature of the GB interactions is the chiral covariant
derivative
\begin{equation}
 D_\mu U = \partial_\mu U - ir_\mu U + iUl_\mu~, ~~ D_\mu U \mapsto RD_\mu U L^\dagger~.
\end{equation}
Here, the left- and right-handed external fields $l_\mu, r_\mu$ are linear combinations of the external
vector and axial-vector-fields $v_\mu,a_\mu$. A second building block encodes the quark masses in the
form
\begin{equation}
\chi = 2 B (s+ip) = 2 B {\cal M} + \ldots~, ~~ B = |\langle 0|\bar{q}q|0\rangle|/F_\pi^2~,
\end{equation} 
where $B$ is a constant with dimension mass and its relation to the scalar quark condensate is derived
below. Further, $\chi$ transforms as $\chi \mapsto R \chi L^\dagger$ under chiral symmetry.
With these building blocks, we are in the position to construct the chiral effective Lagrangian of QCD,
it takes the form 
\begin{equation}
{\cal L}_{\rm eff} = {\cal L}^{(2)} +  {\cal L}^{(4)} + \ldots~,
\end{equation}
where the superscript $(n)$ denotes the {\bf chiral dimension} $d$, that is the number of derivatives or
quark mass insertions, or in terms of a small parameter $q$
\begin{equation}
U, U^\dagger = {\cal O}(p^0) = {\cal O}(1)~, ~~  D_\mu = {\cal O}(p)~, ~~ \chi = {\cal O}(p^2)~,
\end{equation}    
where the power counting of the chiral covariant derivative is natural and  the last relation can be traced
back to the fact that the squared GB masses are proportional to the quark masses, as shown below. Note that
because of parity, we can only have terms with $d$ even.

With this,
it is straightforward to construct the LO Lagrangian ${\cal L}^{(2)}$,
\begin{BoxTypeA}[sec3:box1]{}
\begin{equation}\label{eq:L2}
{\cal L}^{(2)} = \frac{F_\pi^2}{4}\langle D_\mu U D^\mu U^\dagger + \chi U^\dagger + \chi^\dagger U \rangle~,
\end{equation}
\end{BoxTypeA}
\noindent
where $\langle \cdots \rangle$ denotes the trace in flavor space. Using the Noether construction,
one can calculate the axial-current from this Lagrangian and by comparison with Eq.~\eqref{eq:axcurr}
identify the pion decay constant as the overall scale. Clearly, at next-lo-leading order (NLO), we
can have terms with four derivatives, two derivatives and one quark mass insertion or two quark
mass insertions, and correspondingly at higher orders. For the construction of the higher order
terms in the effective Lagrangian, see e.g.~\cite{Gasser:1984gg,Bijnens:1999sh}.

It is most instructive to take a closer look at the mapping of the quark masses onto the GB masses.
For simplicity, let us consider the two-flavor case. To leading order in the quark masses,
the symmetry breaking  Lagrangian can be written as ${\mathcal L}_{\rm SB} = {\mathcal M} \times f(U)$,
${\cal M} = {\rm diag}(m_u,m_d)$ so that at LO, we can can from two invariant combinations from the
$\chi$ and $U$ building blocks, so that
\begin{equation}
{\mathcal L}_{\rm SB} = \frac{1}{2} F_\pi^2  B \, \langle {\cal M}\, (U+U^\dagger)\rangle~,
\end{equation}
with $B$  a real constant if CP is conserved. Expanding now $U = \exp(i \vec\tau\cdot \vec \pi / F_\pi)]$ in powers
of the pion fields leads to
\begin{equation}\label{eq:expand}
{\mathcal L}_{\rm SB} = (m_u+m_d) \, B\, \left[ F_\pi^2 - \frac{1}{2}\pi^2  + \frac{\pi^4}{24F_\pi^2} + \ldots \right]~.
\end{equation}
The first term on the right-hand-side of this equation is related to the vacuum (no pion fields),
and we find upon differentiation with respect to the quark masses
\begin{eqnarray}\label{eq:cond2}
\left.\frac{\partial{\cal L}_{\rm QCD}}{\partial m_q}\right|_{m_q = 0} &=& - \bar q q \nonumber\\
&\Rightarrow& \langle 0 | \bar u u|0\rangle =  \langle 0 | \bar d d|0\rangle = - B F_\pi^2
\left( 1 + {\cal O}{\cal(M)}\right)~,
\end{eqnarray}
which is the sought-after relation between the constant $B$ and the scalar quark condensate. The second term
in the square brackets of Eq.~\eqref{eq:expand} is nothing but the pion mass term
\begin{equation}
-\frac{1}{2}M_\pi^2 \pi^2 \Rightarrow M_\pi^2 = (m_u+m_d) B~,
\end{equation}
which combined with Eq.~\eqref{eq:cond2} gives the GMOR relation
\begin{equation}
M_\pi^2 = -(m_u+m_d)\, \langle 0 | \bar u u|0\rangle /F_\pi^2~.
\end{equation}
What a pleasure to derive this in a few lines, which underlines the efficiency of the effective Lagrangian approach.
In fact, working out the chiral corrections to the GMOR relation, one finds that the this LO term
gives 94\% of the pion mass~\cite{Colangelo:2001sp}.

Repeating this exercise for the three-flavor case, we find at LO:
\begin{equation}\label{eq:massesLO}
M_{\pi^\pm}^2 = (m_u+m_d)B~,~~ M_{K^\pm}^2 = (m_u+m_s)B~,~~ M_{K^0}^2 =  M_{\bar{K}^0}^2 =  (m_d+m_s)B~,
\end{equation}
where we have not given the relations for $M_{\pi^0}$ and $M_\eta$ because these require diagonalization due to the
mixing of the $\Phi^3$ and $\Phi^8$ components, cf. Eq.~\eqref{eq:U}. We can draw a number of conclusions from these relations: First,
it is obvious that $m_s \gg m_u, m_d$ since
\begin{equation}
\frac{m_s}{m_u+m_d} = \frac{M_{K^+}^2 + M_{K^0}^2 - M_{\pi^\pm}^2}{M_{\pi^\pm}^2} \simeq 24~,
\end{equation}  
and second, one obtains the Gell-Mann--Okubo relation $3M_\eta^2 = 4M_K^2 - M_\pi^2$ in terms of the averaged
kaon and pion masses, which is fulfilled within a few percent in Nature. From the relations Eq.~\eqref{eq:massesLO}
one further derives $m_u/m_d = 0.66$, which was already discussed in the context of isospin violation above.
These ratios are of course subject to higher order corrections in the quark masses as well as electromagnetic
effects, but these do not change the conclusions drawn from the LO expressions, see also Sect.~\ref{sec:mq}.

Before proceeding, two notes of caution are in order. In the discussion so far, I have been cavalier with
respect to two issues. First, in the leading order effective Lagrangian, what really appears is the
leading order term of the pion decay constant, that differs from the physical one by terms proportional
to the quark masses. Second, these values are not the same for the two- and the three-flavor cases,
as the strange quark is integrated out in the first case. A similar statement holds for the LEC $B$.
For the precise relations, see e.g.~\cite{Gasser:1984gg}.

\subsection{Loops and power counting}\label{sec:mesonPC}

\begin{figure}[h]
	\centering
        \fbox{\includegraphics[width=.11\textwidth]{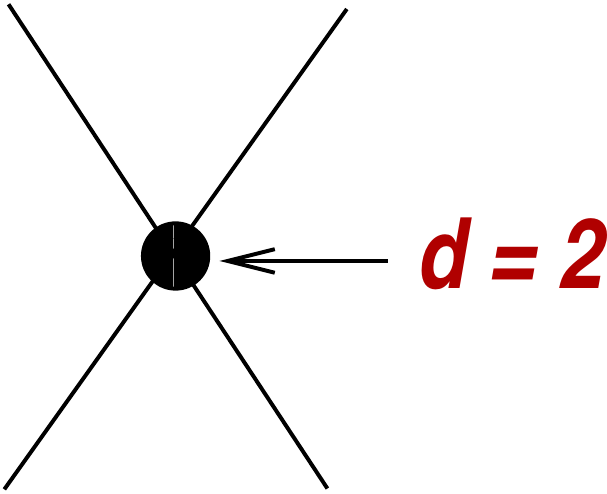}~~~
        \includegraphics[width=.11\textwidth]{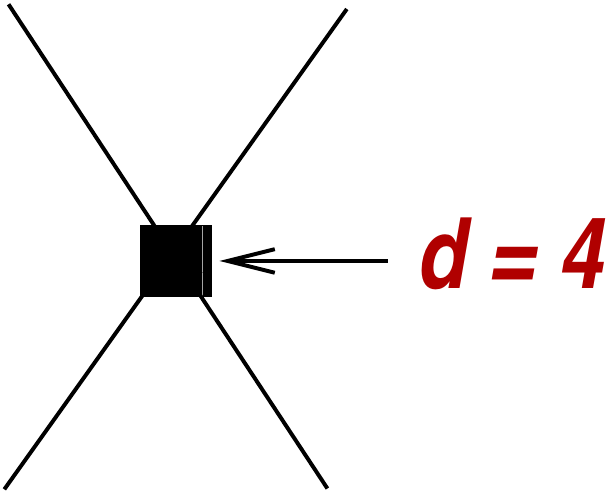}~~~
        \includegraphics[width=.26\textwidth]{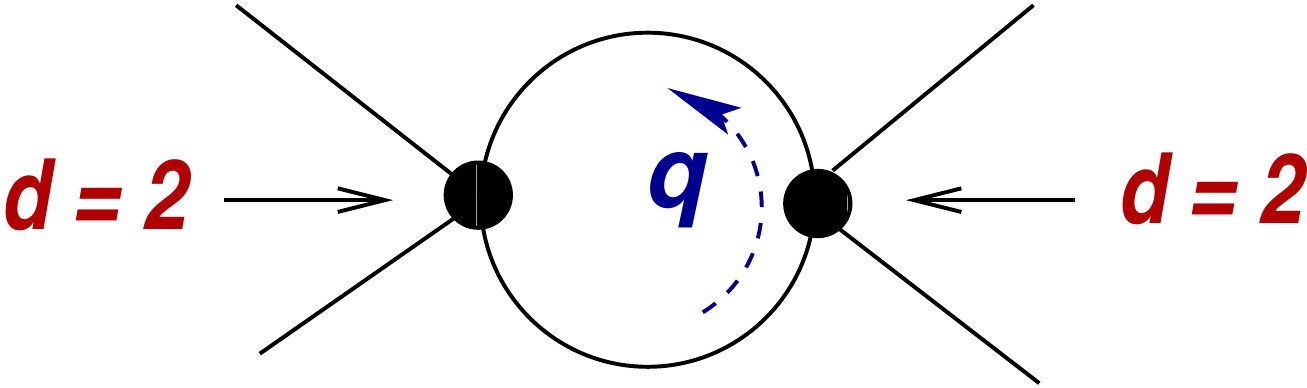}}\hspace{0.7cm}
	\fbox{\includegraphics[width=.15\textwidth]{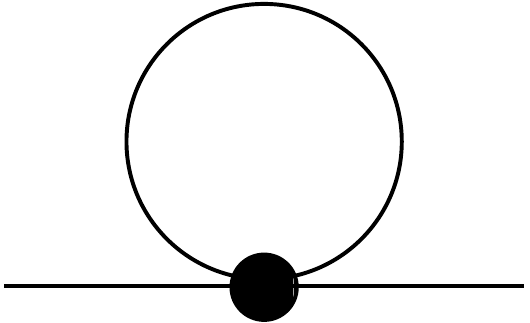}}\hspace{0.7cm}
	\caption{
          Left panel: The pion-pion scattering amplitude
          at one loop (wavefunction and mass renormalization diagrams are not shown).
          From left to right: LO tree graph with one $d=2$ insertion, NLO tree graph
          with one $d=4$ insertion and a NLO loop diagram build entirely from  $d=2$ operators.
          Right panel: The pion tadpole diagram at one loop.}
	\label{fig:loops}
\end{figure}

The expansion in small momenta and masses can be mapped onto a loop expansion, as can be understood
most easily by patching together two LO interactions graphs with chiral dimension $d=2$, as shown in
in the left panel of Fig.~\ref{fig:loops}. Further, loops generate imaginary parts, which is
important, as scattering amplitudes or form factors are complex-valued in Nature. I will come back to this issue below.
It is important to realize that loop diagrams are in general divergent. Naive dimensional counting
of the $\pi\pi\to\pi\pi$  loop diagram in the left panel of Fig.~\ref{fig:loops} gives
\begin{equation}
Amp \sim  \int d^4q \frac{q_1 \cdot q_2 \,\, q_3\cdot q_4}{(q^2-M_\pi^2)(q^2 - M_\pi^2)} \sim {\mathcal O}(q^4)
\end{equation}
as the LO pion-pion interaction vertex involves two momenta and $q_i$ is the momentum of pion
$i$ ($i=1,2,3,4)$. Clearly, this diagram is ultraviolet (UV) divergent,
if we e.g. utilize a momentum cutoff $\Lambda$, it scales as $\Lambda^4$. Let us make this more explicit by
calculating the simpler tadpole diagram shown in the right panel of Fig.~\ref{fig:loops}. Let us choose a
mass-independent and symmetry-preserving regularization scheme like dimensional regularization (dim. reg.)
\begin{eqnarray}
-i \Delta_\pi (0) &=& \frac{-i}{(2\pi)^d}
\int d^dp \frac{1}{M^2 - p^2 -i\varepsilon} =(2\pi)^{-d} \int d^dk \frac{1}{M^2+k^2}~{\rm with}~~p_0 = ik_0~,~~
-p^2 = k_0^2 + {\vec k\,}^2 \nonumber\\
&=& (2\pi)^{-d} \int d^dk \int_0^\infty d\lambda \exp(-\lambda (M^2+k^2))
= (2\pi)^{-d}  \int_0^\infty d\lambda \exp(-\lambda M^2) 
\underbrace{\int d^d k\exp(-\lambda k^2)}_{(\pi / \lambda)^{d/2}}\nonumber\\[-1.9ex]
&=& (4\pi)^{-d} \, M^{d-2} \, \Gamma\left(1 -\frac{d}{2}\right)~,
\end{eqnarray}  
where $\Gamma(1-d/2)$ has a pole at $d=4$. Since the effective Lagrangian contains all terms consistent with the symmetries,
it is always possible to absorb these divergences in the LECs at next-to-leading order (for one-loop
diagrams), symbolically written as
\begin{equation}
g_i \to g_i^{\rm ren} + \beta_i \frac{1}{d-4}~,
\end{equation}  
where $\beta_i$ is a calculable number and the finite value $ g_i^{\rm ren}$ must be determined by a fit to data.

At this point, let  us make some general remarks on the LECs. For that, consider a covariant and
parity-invariant theory of Goldstone bosons parameterized in some matrix-valued field $U$,
\begin{equation}
{\cal L}_{\rm eff} = g_2 \,\langle\partial_\mu U \partial^\mu U^\dagger\rangle
+ g_4^{(1)} \,\langle \partial_\mu U \partial^\mu U^\dagger\rangle^2
+ g_4^{(2)} \,\langle \partial_\mu U \partial^\nu U^\dagger\rangle
\langle \partial_\nu U \partial^\mu U^\dagger\rangle + \ldots~,
\end{equation}
where the appearing couplings $g_d^{(n)}$ are the LECs. As we already know, $g_2 \neq 0$ because of
spontaneous chiral symmetry breaking and the LECs $g_4^{(1)}, g_4^{(2)}, \ldots$ must be fixed from data (or
be  calculated from the underlying theory). Calculations in CHPT thus are done in two steps, first one
fixes the LECs from some processes and then makes predictions for other reactions. It is also important to stress that the
LECs encode information from the high mass states that are integrated out, which leads to the concept
of {\bf resonance saturation}, which is most easily understood by considering $\rho$-meson exchange
in $\pi\pi$ scattering, $\pi\pi\to\rho\to\pi\pi$,
\begin{equation}
T \approx \frac{g_{\rho\pi\pi}^2 q^2}{M_\rho^2-q^2}  \stackrel{q^2\ll M_\rho^2}{\longrightarrow}
\frac{g_{\rho\pi\pi}^2 q^2}{M_\rho^2}\left(1+ \frac{q^2}{M_\rho^2} + \ldots \right)~,
\end{equation}  
suppressing all Dirac structures and other indices, leading to a tower of four-pion contact terms with
even number of derivatives and coupling strengths given in  terms of the $\rho\pi\pi$ coupling constant $g_{\rho\pi\pi}$
and inverse powers of the $\rho$ mass $M_\rho$. For more general discussions on this,
see~\cite{Donoghue:1988ed,Ecker:1988te}.

Having elucidated the appearance of UV divergences in loop diagrams, we can now work out the {\bf power
counting} that allows us to collect all terms at a given order. Consider ${\cal L}_{\rm eff} = \sum_d {\cal L}^{(d)}$
with $d$ bounded from below. For interacting Goldstone bosons, $d \ge 2$, and the propagator is $iD(q) = {1}/(q^2-M^2)$,
with $M^2$ the leading term in the quark mass expansion of the pion mass. Consider an $L$-loop diagram with $I$ internal
lines and $V_d$ vertices of order $d$,
\begin{equation}\label{eq:amppcmeson}
Amp \propto \int (d^4q)^L \, \displaystyle\frac{1}{(q^2)^I} \prod\limits_d (q^d)^{V_d}~.
\end{equation}  
Now,  let $Amp \sim q^\nu$ so that $\nu = 4L -2I + \sum\limits_d d V_d$, use the topological relation
$L = I - \sum_d V_d + 1$ and eliminate $I$. One obtains
\begin{BoxTypeA}[sec3:box2]{}
\begin{equation}\label{eq:pcmeson}
\nu = 2 + 2L + \sum\limits_d V_d (d-2)~.
\end{equation}
\end{BoxTypeA}
\noindent
This gives the structure already displayed in Fig.~\ref{fig:loops} for the pion-pion scattering amplitude.
At LO, we have tree level graphs with one $d=2$ insertion. At NLO, there are tree diagrams with one $d=4$ insertion
as well as one-loop diagrams with $d=2$ insertions only, that scale with the fourth power of the chiral dimension.
The appearing UV
divergences are renormalized by the corresponding $d=4$ LECs. This nicely exhibits that the EFT is
renormalizable order-by-order, provided a proper regularization scheme is employed (as dim. reg. in the
above example). More generally, from Eq.~\eqref{eq:pcmeson} it follows that there is a one-to-one
correspondence between the loop expansion and the one in small momenta/masses, that is an $N$-loop
diagram scales as ${\cal O}(q^{2+2N})$~($N=0,1,2,\ldots$)\,.

As stated above, loop diagrams generate imaginary parts, this can be easily seen by applying the Cutkovsky
cutting rules to the loop diagram shown in Fig.~\ref{fig:loops}, see e.g.~\cite{Itzykson:1980rh}. The imaginary
part of the $\pi\pi$ scattering amplitude at one-loop (NLO), $T = T^{(2)}+T^{(4)}$, can be worked out easily,
\begin{equation}
  {\rm Im}~T^{(4)} = \frac{1}{16 \pi F_\pi^4} \sqrt{1-\frac{4M_\pi^2}{s}} (s-M_\pi^2)
  = \frac{\sigma}{16\pi}\, (T^{(2)})^2~,
\end{equation}  
in terms of the LO scattering amplitude $T^{(2)}(s,t,u) = (s-M_\pi^2)/F_\pi^2$, with $s,t,u$ the conventional
Mandelstam variables and, in particular, $s$ is the center-of-mass energy squared. This clearly shows that
the unitary part is only non-vanishing above the two-pion threshold (as it should) and that unitarity is
restored perturbatively in CHPT.

\subsection{Chiral anomalies and the Wess-Zumino-Witten term}

Here, I present a short discussion of the anomalous sector of QCD and how it appears in CHPT.
For further details, see e.g.~\cite{Wess:1971yu,Witten:1983tw,Gasser:1984gg,Bertlmann:1996xk}.
Considering the low-energy theory of QCD,
we note  at LO the chiral effective Lagrangian Eq.~\eqref{eq:L2}, which is nothing but the time-honored
non-linear $\sigma$-model, exhibits redundant symmetries.
For simplicity, let us consider the massless case,
\begin{equation}\label{eq:NLS}
 {\cal L}^{(2)} = \frac{F_\pi^2}{4} \langle \partial_\mu U \partial^\mu U^\dagger\rangle~,~~
U(x) =\exp\left(i\lambda^a \Phi^a(x)/F_\pi\right)~,~U\in {\rm SU}(3) 
\end{equation}
which  is obviously invariant under parity, $PU(\vec{x},t)P^{-1} = U^\dagger(-\vec{x},t)$, just like QCD.
However, ${\cal L}^{(2)}$ has two extra symmetries, unlike QCD, as the transformations
\begin{equation}
U(\vec{x},t) \mapsto U(-\vec{x},t)~,~~~~U(\vec{x},t) \mapsto U^\dagger(\vec{x},t)~,
\end{equation}  
conserve intrinsic parity ($P_I$), where $P_I = +1/-1$ for a true/pseudo-tensor of rank $k$.
Consider a few examples:
\begin{eqnarray}\label{eq:anop}
&&\pi\pi \to \pi \pi:~~~(-1)\cdot(-1) = (-1)\cdot(-1)~~\surd
~~~~~~~~~~~~~~~~ \pi^0\to 2\gamma:~~~(-1) = (+1)\cdot(+1)~~? \nonumber\\
&&\gamma\pi^+ \to \pi^+:~~~(+1)\cdot(-1) = (-1)~~\surd
~~~~~~~~~~~~~~~~~~~~~~~~ \gamma\to\pi^+\pi^-\pi^0:~~~(+1) = (-1)\cdot(-1)\cdot(-1)~~?  \nonumber\\
&& \eta\to\pi^+\pi^-\pi^0:~~~(-1) = (-1)\cdot(-1)\cdot(-1)~~\surd
~~~~~~~~~~ K\bar{K}\to \pi^+\pi^-\pi^0~~~(-1)^2 = (-1)^3~~?~.  
\end{eqnarray}  
Thus,  ${\cal L}^{(2)}$ conserves $P_I$, i.e. the number of Goldstone bosons mod 2
[in fact, this holds for all ${\cal L}^{(2n)}$]  as denoted by the symbol $\surd$, but QCD does not,
shown by the $?$ in the equations above. For instance, the decay $2K\to3\pi$ is described in effective
chiral models by a vertex $\epsilon_{\mu\nu\alpha\beta}\pi^0\partial^\mu K^+ \partial^\nu K^- \partial^\alpha \pi^+
\partial^\beta \pi^-$~\cite{Wess:1971yu}, which clearly breaks intrinsic parity. In the QCD language, this
corresponds to the $(AAAAA)$ pentagon anomaly~\cite{Bardeen:1969md}.
In fact, all the decay processes in Eq.~\eqref{eq:anop}
that violate intrinsic parity are mediated by anomalies in QCD. This information is not included
in the non-linear $\sigma$-model~\eqref{eq:NLS}, since the resulting Ward identities are anomaly-free.
To account for this, Witten~\cite{Witten:1983tw} proposed modifying the equations of motion associated
to~\eqref{eq:NLS} by adding a chiral invariant term that breaks the redundant symmetries, while preserving
parity
\begin{equation}\label{eq:EoM}
\frac{i}{2}F_\pi^2 \partial^\mu L_\mu + \lambda \epsilon^{\mu\nu\alpha\beta}
     L_\mu L_\nu L_\alpha L_\beta + \cdots = 0~,~~~~~L_\mu = U^\dagger \partial_\mu U~,
\end{equation}
where the dots stand for higher order terms. Such a term would make the processes in Eq.~\eqref{eq:anop}
marked with a ? possible.  Unfortunately, it is not possible to integrate the equations of motion~\eqref{eq:EoM}
to a manifestly chiral invariant action in four dimensions. Using general topological arguments, it can be
shown that the second term in Eq.~\eqref{eq:EoM} is related to a closed topological invariant in 5-dimensional
space that is manifestly chiral invariant. For that, $U(\vec{x},t)$ in $\mathbb{R}^4$ is extended to $U(\vec{x},t,s)$,
so that $U: M_5=\mathbb{R}^4\times[0,1]\to S^5$. Realizing that this term is topologically quantized,
that is $\lambda = -in/(48\pi^2)$, with $n \in \mathbb{Z}$, it follows that  the Wess-Zumino-Witten (WZW)
action takes the form (a detailed derivation is given e.g. in~\cite{Meissner:1985qs})
\begin{BoxTypeA}[sec3:box3]{}
\begin{equation}\label{eq:WZW}
S_{\rm WZW} = -\frac{i n}{240\pi^2} \int_{S^5} d^5x \, \epsilon^{\mu\nu\alpha\beta}\,
\langle  L_\mu L_\nu L_\alpha L_\beta L_\gamma\rangle~,
\end{equation}
\end{BoxTypeA}
\noindent
which embodies all chiral QCD anomalies.  Electromagnetic gauging leads to $n=N_C$, e.g. via the
process $\pi^0\to 2\gamma$. The WZW action embodies many testable predictions, like for $\gamma\to \pi^+\pi^-\pi^0$.
This will be discussed later. Note further that this term appears at NLO, that is ${\cal O}(q^4)$. For
a more detailed derivation in the CHPT framework, see~\cite{Gasser:1984gg}.

\subsection{The chiral limit}

The chiral limit of QCD shows some peculiarities, which deserve a short discussion. In what follows,
I will consider the two-flavor case for simplicity, but all statements made can be easily transferred
to three flavors. In this limit, the GBs (the pions) are massless, so that the  pion propagator (in
the static approximation) becomes Yukawa-like:
\begin{equation}\label{eq:tail}
G(\vec{r}-\vec{r}^\prime) = \frac{1}{4\pi}\frac{e^{-M_\pi |\vec{r}-\vec{r}^\prime|}}{|\vec{r}-\vec{r}^\prime|}~~
\stackrel{M_\pi \to 0}{\longrightarrow}~~\frac{1}{4\pi}\frac{1}{|\vec{r}-\vec{r}^\prime|}~.
\end{equation}  
This explains why certain quantities, that exhibit so-called ``chiral logarithms'', diverge in the
chiral limit. Consider e.g. the pion charge radius, that contains such a chiral log~\cite{Beg:1972vyx}:
\begin{equation}
\langle r^2\rangle_V^\pi = -\frac{1}{16\pi^2F_\pi^2}\ln M_\pi^2 + \ldots~.
\end{equation}   
It is physically clear why such a quantity does not exist in the chiral limit: The pion cloud surrounding
any particle, in particular the cloud surrounding the pion, becomes long-ranged as $M_\pi\to 0$, as there
is no more Yukawa suppression to cut it off a large distances. In the chiral limit the charge distribution
falls off with a power of the distance, see Eq.~\eqref{eq:tail}, and the mean square radius of the distribution
diverges. Other examples of this type of behavior are the scalar pion radius or the nucleon isovector
electromagnetic radii (where the lowest-mass contribution to the spectral functions is given by two-pion
exchange). Not surprisingly, certain two-photon observables like the pion and the nucleon polarizabilities
diverge even like $1/M_\pi$ in the chiral limit~\cite{Babusci:1991sk,Bernard:1991rq}. From this we
conclude that the approach to the chiral limit is non-analytic in the quark masses, as witnessed by the
chiral logs. All this is encapsulated in the following decoupling theorem: The leading chiral non-analytic
terms stem from pion (GB) one-loop graphs coupled to pions (GBs) or nucleons (ground state baryons)~\cite{Gasser:1979hf}.
This needs to be accounted for if one extends the EFT to include resonance fields like the $\rho$ or the
$\Delta$. In particular, it can also be shown that the chiral limit of QCD and the large-$N_C$ limit
do not commute. However, S-matrix elements exist in the chiral limit for arbitrary momenta~\cite{Langacker:1973hh}.

Moving away from the chiral limit also determines the power counting discussed above.  
Indeed, the dominating feature in the low-energy region is the occurrence of the GBs.
In the chiral limit the Green functions develop poles at $p^2=0$ and cuts starting at $p^2=0$.
If the masses of the light quarks are switched on, the poles move to $p^2=M_\pi^2= {\cal O}({\cal M})$
and the cuts start at $4M_\pi^2$, $9M_\pi^2$ and so on. To properly 
describe the structure of these low-energy singularities, one has to treat the momenta as
quantities of the same algebraic order as $M_\pi$,, i.e., consider an expansion in powers of the
momenta and of the quark masses at fixed ratio $M_\pi^2/p^2\sim {\cal M}/p^2$.

Finally, we note  that in a finite volume ($V=L^3$) or at finite temperature $T$, a new
energy scale $\sim 1/L$ or $T$, respectively, appears. This leads to a more complicated
power counting related to the relative size of the scales $M_\pi,T$ and $1/L$~\cite{Leutwyler:1987ak}.
In particular, in a finite box the chiral limit mass of the pion does not vanish but is bounded
from below. For a more detailed discussion of these topics, see e.g. Ref.~\cite{Meissner:2022cbi}.

\subsection{Low-energy  theorems}

In the realm of quantum field theory, low-energy theorems (LETs) have first been considered in
Quantum Electrodynamics (QED).  For the scattering of very soft photons on the proton,
using gauge invariance and the fact that the $T$-matrix can be written in terms of a
time-ordered product of two conserved vector currents sandwiched between
proton states,  one arrives at LETs for the spin-independent and the spin-dependent
scattering amplitudes $T_0$ and $T_1$, respectively, for vanishing photon momenta
\cite{Low:1954kd,Gell-Mann:1954wra}
\begin{equation}
T_0 = - \frac{Z_p^2e^2}{4 \pi m_p} \vec{\varepsilon}\, ' \cdot \vec{\varepsilon}~, \quad
T_1 = - \frac{Z_p^2e^2 \kappa_p^2}{8 \pi m_p} \frac{|\vec{k}_1|}{m_p}\vec{\sigma} \cdot (
\vec{\varepsilon}\, ' \times \vec{\varepsilon})~,
\end{equation}
with $\vec{k}_1$ the momentum of the incoming photon, $m_p$ the proton mass, $Z_p=1$ its charge
and $\kappa_p = 1.793$ its anomalous magnetic moment.  These LETs are physically intuitive, as
very soft photons have a very low resolution, so that they are only sensitive to bulk properties of the
proton, like its mass and its magnetic moment. Such LETs have been derived for the strong interactions
even before the formulation of QCD and its low-energy EFT, namely CHPT. The broken chiral symmetries
of the strong interactions can be analyzed in terms of its (partially) conserved currents and the  almost massless
pion plays the role of the soft photons in the above QED example, so one was also talking about ``soft-pion
theorems''. Arguably the most prominent example is Weinberg's work on pion scattering lengths~\cite{Weinberg:1966kf},
which leads to the LO $\pi\pi$ scattering lengths discussed in Sect.~\ref{sec:pipi} and the predictions for
the S-wave pion-nucleon scattering lengths, 
\begin{equation}\label{eq:WpiN}
a_{\pi N}^+ = 0~, \quad  a_{\pi N}^- = \frac{M_\pi}{8\pi F_\pi^2}\frac{1}{1+M_\pi/m_p} = 0.079 M_\pi^{-1}~,  
\end{equation}
where the superscripts $+/-$ refer to the isoscalar and the isovector $\pi N$ scattering amplitude, in order.
One can confront these LETs with data (see e.g.~\cite{Hoferichter:2015hva}), but is should be said that
with the advent of CHPT, a more general definition a LET became available (see e.g. the discussion of the
scalar radius of the pion in Ref.~\cite{Gasser:1983yg}): 
\begin{quote}
	\quotehead{Definition of a LET}
	A Low Energy Theorem (LET) of order ${\cal O}(q^n)$ is a general prediction of CHPT to ${\cal O}(q^n)$.
	\source{Ecker and Mei{\ss}ner~\cite{Ecker:1994ra}}
\end{quote}
As will be shown below, this is not an purely academic exercise but it really lead to a deeper understanding
of certain processes in chiral dynamics.

\subsection{Assorted applications}\label{sec:mesonapp}

Here, I will discuss a few selected applications and refer to a number of reviews for many more
interesting and detailed discussions of CHPT and its applications to GB physics
\cite{Meissner:1993ah,Bijnens:1993xi,Ecker:1994gg,Pich:1995bw,Bernard:2006gx,Bijnens:2006zp}.

\subsubsection{Pion-pion scattering}
\label{sec:pipi}

Arguably the purest process in two-flavor CHPT is pion-pion scattering, as it only involves the
light up and down quarks. Consider the process $\pi^a(p_a)+\pi^b(p_b)\to\pi^c(p_c)+\pi^d(p_d)$
in terms of the Mandelstam variables $s=W^2=(p_a+p_b)^2=(p_c+p_d)^2$, $t = (p_a-p_c)^2 = (p_d-p_b)^2
=-2q^2(1-\cos \theta)$, $u = (p_a-p_d)^2 = (p_d-p_b)^2 =-2q^2(1+\cos \theta)$, subject to the constraint
$s+t+u=4M_\pi^2$, with $W$ the center-of-mass (CMS) energy, $q$ the exchanged momentum and $\theta$
the scattering angle in the CMS. If not noted otherwise, $M_\pi$ is given by the charged pion mass.
The scattering amplitude
\begin{equation}
T^{ab;cd}(s,t,u) = A(s,t,u) \delta^{ab} \delta^{cb} + A(t,s,u) \delta^{ac} \delta^{bd} + A(u,t,s) \delta^{ad} \delta^{bc}~, 
\end{equation}
is given in terms of one single function $A(s,t,u)$. As the pion has isospin one, the $\pi\pi$ scattering
amplitudes with definite isospin ($I=0,1,2$) follow as
\begin{equation}
T^0 = 3A(s,t,u)+A(t,s,u)+A(u,t,s)~,~~~T^1 = A(t,s,u)-A(u,t,s)~,~~~T^2 = A(t,s,u)+A(u,t,s)~.
\end{equation}  
Projection on partial waves with definite isospin $I$ and definite angular momentum $\ell$ is given by
\begin{equation}
T_\ell^I(s) = \frac{1}{64\pi}\int_{-1}^{+1}dx \, P_\ell (x)\,T^I(s,t,u)~, ~~~~ x = \cos\theta~,
\end{equation}
with $P_\ell$ the Legendre polynomials. In the elastic region $4M_\pi^2<s<16M_\pi^2$, the partial 
wave amplitudes $t_\ell^I$ are described by real phase shifts $\delta_\ell^I$, 
\begin{equation}
t_\ell^I(s) = \left( \frac{s}{s-4M_\pi^2} \right)^2\, \frac{1}{2i}\,\left\{e^{2i \delta_\ell^I(s)}-1 \right\}~.
\end{equation}
Furthermore, in the threshold region the $t_\ell^I(s)$ obey the effective range expansion,
\begin{equation}
{\rm Re}~t_\ell^I(s) = q^{2\ell}\,\left(a_\ell^I + q^2\,b_\ell^I + \ldots\right)~,~~I=0,1,2~,~~~\ell = 0,1,2,...~, 
\end{equation}
in terms of the scattering lengths $a_\ell^I$ and effective ranges $b_\ell^I$ and one commonly uses the
spectroscopic notation: $\ell=0$ corresponds to the S-wave, $\ell=1$ to the P-wave and so on. In what follows,
I will focus on the two S-waves scattering lengths, that is $a_0^0$ and $a_0^2$, because these serve
as precision  tests of the chiral structure of QCD~\cite{Gasser:1983kx}. From the LO Lagrangian,
one can easily derive the expressions for these~\cite{Weinberg:1966kf} 
\begin{equation}
a_0^0 = \frac{7 M_\pi^2}{32 \pi F_\pi^2}~, ~~~~~ a_0^2 = -\frac{M_\pi^2}{16 \pi F_\pi^2}~.
\end{equation}
Note that in the chiral limit of vanishing pion mass, these scattering lengths also vanish, they
are thus very sensitive to the precise pattern of explicit chiral symmetry breaking.
The corrections at NLO (one loop) and NNLO (two loops) have been worked out in~\cite{Gasser:1983kx}
and~\cite{Bijnens:1995yn}, respectively, with the following results:
\begin{eqnarray}\label{eq:a0}
  {\rm LO} &:& ~~~~ a_0^0 = 0.16~, ~~~~~~~~~~~~~~~~~~~~a_0^2 = -0.045~ \nonumber\\
  {\rm NLO} &:& ~~~~  a_0^0 = 0.20\pm 0.01~, ~~~~~~~~ a_0^2 = -0.042\pm 0.002~ \nonumber\\
  {\rm NNLO} &:& ~~~~ a_0^0 = 0.217\pm 0.009~, ~~~~ a_0^2 = -0.0413\pm 0.0030~,
\end{eqnarray}
where in case for the two-loop result, the central value is obtained from set~I of the LECs
in~\cite{Bijnens:1997vq} and the uncertainty from the difference to set~II of the LECs.
These results display some intriguing trends. For the $I=2$ case, the corrections to the LO
results are small, consistent with the expansion parameter $(M_\pi/4\pi F_\pi)^2 = 0.0145$.
Matters are very different for the isospin zero case, the correction from LO to NLO is about
$25\%$ and from NLO to NNLO still $8.5\%$. These are much larger than expected on dimensional grounds.
Combining the two-loop CHPT amplitude with dispersion relations (Roy equations), one can even sharpen the
prediction for the S-wave scattering lengths~\cite{Colangelo:2000jc},
\begin{equation}
{\rm Roy~eq. + CHPT}: ~~~~ a_0^0 = 0.220\pm0.001~, ~~~~~~ a_0^2=-0.0444\pm0.0010~,
\end{equation}
which are truly remarkable predictions in low-energy QCD. These align
well with the experimental determinations based on $\pi\pi$ final-state interactions in kaon
decays~\cite{NA482:2010dug},
\begin{equation}
a_0^0 =  0.2210 \pm 0.0047_{\rm stat.} \pm 0.0040_{\rm syst.}~, \quad
a_0^2 = -0.0429 \pm 0.0044_{\rm stat.} \pm 0.0028_{\rm syst.}~.
\end{equation}  

\begin{figure}[t]
	\centering
        \includegraphics[width=.66\textwidth]{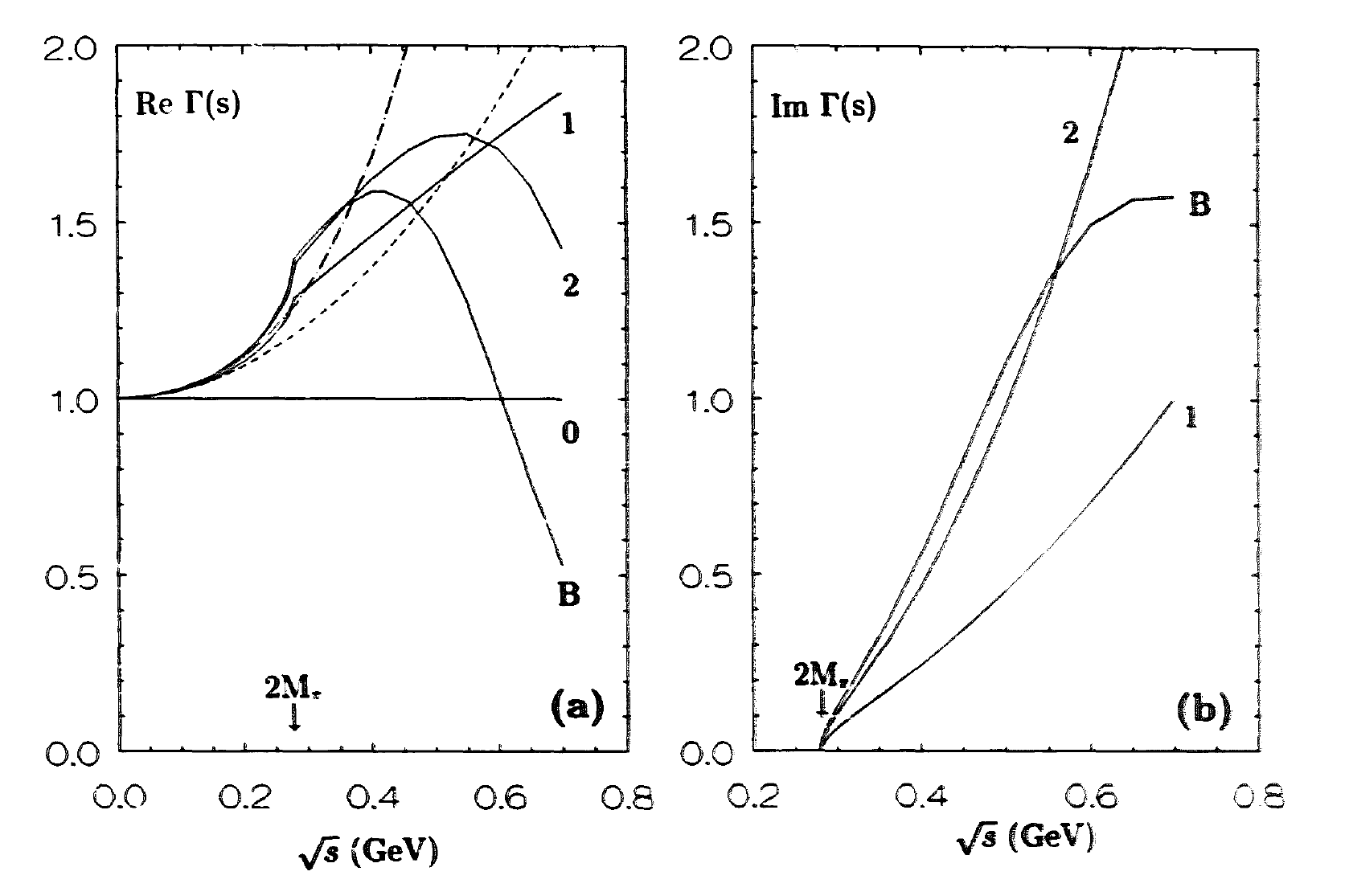}
        \caption{The scalar form factor of the pion, real (left panel) and imaginary (right panel) part.
          The lines labeled by 0,1,and 2 show the result of the tree level, one- and two-loop calculation,
          respectively. Also shown are the polynomial contributions at one loop (dashed line) and at two loop
          (dot-dashed line). The dispersive solution (B) from Ref.~\cite{Au:1986vs} is also given.
          Figure taken from~\cite{Gasser:1990bv}.}
	\label{fig:sff}
\end{figure}

It was long speculated that the strong final-state interaction (FSI) in the isospin-zero S-wave are
due to a low-lying scalar meson, see e.g. the discussion in~\cite{Meissner:1990kz}. To understand
the strong FSI in CHPT, consider the scalar form factor of the pion~\cite{Gasser:1990bv}
\begin{equation}
  \langle \pi^a(p')\pi^b(p)~{\rm out}|\hat{m} (\bar{u}u+\bar{d}d)|0\rangle = \delta^{ab}\, \sigma_\pi \,\Gamma(s)~,
  ~~~~~ s = (p'+p)^2~, ~~~~~ \hat{m} = (m_u+m_d)/2~,
\end{equation}
where for convenience the pion sigma-term ($\sigma_\pi$) has been scaled out, so that $\Gamma(0)=1$.
This is the most basic and simplest object to study the strong FSI.  The scalar form factor is
not directly accessible to experiment, but it can be evaluated from the available $\pi\pi$ scattering data
by means of a dispersive analysis, the so-called Omn\`es-Muskhelisihvili problem~\cite{Donoghue:1990xh} (and
references therein). As can been seen from Fig.~\ref{fig:sff}, the one-loop representation in the real
part already exhibits sizeable corrections to the tree level result at the two-pion threshold (about 30\%)
and the imaginary part very quickly deviates from the
dispersive result from~\cite{Au:1986vs}. Things are different at two-loop order, the real part now shows
the required bending but starts to overshoot the dispersive curve at $\sqrt{s}\simeq 0.5\,$GeV, and
deviations of the two-loop to the dispersive result behave similarly for the imaginary part. Clearly,
the two-loop representation captures the essence
of the strong FSI. It is instructive to consider the polynomial terms in the chiral expansion, these
read $1+ c_1s$ and $1+c_1s+c_2s^2$ at one- and two-loop order, with the $c_i$ known constants, see the
dashed and the dot-dashed lines in the left panel of Fig.~\ref{fig:sff}, respectively.  One
sees that the strong FSI manifest themselves in large unitarity corrections, that also lead to
the large corrections exhibited in Eq.~\eqref{eq:a0}.
Note further that the shape of the real part above the two-pion threshold is reminiscent
of a broad scalar meson, as detailed in~\cite{Meissner:1990kz}.
We know now that the lowest resonance in QCD, the $f_0(500)$ meson, is located deep inside the
complex plane, $M_\sigma = 441^{+16}_{-8}\,$MeV and  $\Gamma_\sigma =  272^{+9}_{-13}\,$MeV~\cite{Caprini:2005zr},
that reflects itself in the particular shape of the scalar form factor. Quite different from this,
in the P-wave, that features the $\rho$ resonance, the two-loop corrections are much more modest,
and both the one-loop and the two-loop result are dominated by the polynomial terms~\cite{Gasser:1990bv},
see also~\cite{Bijnens:1998fm}. This can
be understood form the vector-meson dominance formula $F_V^{\rm VMD} = 1/(1-s/\bar{m}_\rho^2)$ with
the effective $\rho$-meson $\bar{m}_\rho$ adjusted so as to reproduce the vector radius of the
pion, $\bar{m}_\rho^{-2} = \langle r^2\rangle_V^2/6$ and $\langle r^2\rangle_V^2
\simeq (0.44\,{\rm fm})^2$. In this case, the unitarity corrections are very small.

\subsubsection{Electromagnetic corrections to the quark mass ratios}
\label{sec:mq}

From Eq.~\eqref{eq:massesLO}, we can deduce the quark mass ratios at LO, namely
\begin{equation}\label{eq:mqratiosLO}
\frac{m_u}{m_d} = 0.66~, ~~~~ \frac{m_s}{m_d} = 20.1~, ~~~~ \frac{m_u+m_d}{m_s} = \frac{1}{24.1}~.
\end{equation}  
Here, I want to discuss briefly the first corrections due to photons loops, which affect the
charged Goldstone bosons, that the is the $\pi^\pm$ and $K^\pm$. To leading order in
electromagnetism, we can write the mass of a GB as
\begin{equation}
M_P^2 = \left( M_P^2\right)_{\rm QCD} + e^2 Q_P \tilde{C} + {\cal O}(e^2 {\cal M})~, 
\end{equation}  
where $Q_P$ is the charge of the pseudoscalar meson under consideration and the task
is to determine the LEC $\tilde{C}$. One way is to use sum rule techniques~\cite{Das:1967it}
or to resort to Dashen's theorem~\cite{Dashen:1969eg} that states
that the squared mass differences between the charged pseudoscalar mesons
and their corresponding neutral partners are equal in the chiral limit: $M_{K^+}^2 - M_{K^0}^2 =
M_{\pi^+}^2 - M_{\pi^0}^2$. Using the pion mass difference, that is affected by strong
corrections of order ${\cal O}((m_d-m_u)^2)$ only, we find $\tilde{C} = 1.38\cdot10^{-2}\,$GeV$^2$.
This leads to  the famous Weinberg ratios from 1977~\cite{Weinberg:1977hb}
\begin{equation}
\frac{m_d}{m_u} = \frac{M_{K^0}^2 - M_{K^+}^2 + M_{\pi^+}^2}{M_{K^+}^2 - M_{K^0}^2 + 2M_{\pi^0}^2 - M_{\pi^+}^2}~,~~~~
\frac{m_s}{m_d} = \frac{M_{K^0}^2 + M_{K^+}^2 - M_{\pi^+}^2}{M_{K^0}^2 - M_{K^+}^2 + M_{\pi^+}^2}~ 
\end{equation}  
leading to
\begin{equation}
\frac{m_u}{m_d} = 0.55~, ~~~~  \frac{m_d}{m_s} = \frac{1}{20.1}~, ~~~~ \frac{m_u+m_d}{m_s} = \frac{1}{25.9}~, 
\end{equation} 
which differ by only a few percent from the LO results Eq.~\eqref{eq:mqratiosLO}. In fact, these
were predated by Gasser and Leutwyler in 1975~\cite{Gasser:1974wd} (though in a much less transparent
way). Of course, within CHPT, this problem has been tackled to higher precision, based on the
chiral Lagrangian supplemented with virtual photons, see e.g. Refs.~\cite{Urech:1994hd,Knecht:1997jw,Meissner:1997fa}
and for the early  calculations of the corrections to Dashen's theorem, see~\cite{Bijnens:1993ae,Donoghue:1993hj}.

There is one issue that deserves more attention. In the above, we have assumed that we can neatly separate
the QED from the QCD effects. In fact, when the electromagnetic interactions are switched off in QCD+QED, an inherent ambiguity arises,
which is not present in the case of purely strong isospin-breaking due to $m_u\neq m_d$~\cite{Gasser:2003hk}.
This concept of ``switching off'' the electromagnetic interactions is scale-dependent, because the parameters run with different
renormalization group equations when $e=0$ and $e\neq 0$. One way to proceed is to define some matching scale $\mu_1$ to separate the
QED and QCD effects, but this in turn implies that the QCD parameters $g$ and $m_q$ apart from depending on the running scale $\mu$
implicitly also depend on $\mu_1$, leading to a convention-dependence, as witnessed by the different schemes used in lattice
calculations, see e.g.~\cite{FlavourLatticeAveragingGroupFLAG:2021npn}.
One thus has to be careful in comparing results on isospin-breaking effects that have been obtained using different prescriptions
for separating QED from QCD.

\subsubsection{Tests of the chiral anomaly}

The issue of anomalies in fact predates QCD and CHPT. In the pioneering work of Adler, Bell and Jackiw
\cite{Adler:1969gk,Bell:1969ts} it was found that the triangle graphs of the type $(AVV)$, where $A$ is
represented by the derivative of the pion field and $V$ by the photon field, leads to a finite neutral pion
lifetime,
\begin{equation}
\Gamma_{\pi^0\gamma\gamma}= \frac{\alpha^2 M_\pi^3}{64\pi^3 F_\pi^2} = 7.76~{\rm eV}~.
\end{equation}
In modern language, this LO prediction can also be derived from the WZW action~\eqref{eq:WZW} (as noted,
it is used to pin down the constant $n$ in the topological quantization condition).
More interesting are indeed the isospin breaking corrections, that have been worked out to one- and even
two-loop accuracy~\cite{Ananthanarayan:2002kj,Goity:2002nn,Kampf:2009tk}:
\begin{eqnarray}
{\rm one-loop}:~~\Gamma_{\pi^0\gamma\gamma} &=& 8.06\pm 0.02\pm 0.06~{\rm eV}~,\nonumber\\
&=& 8.10 \pm 0.08~{\rm eV}~, \nonumber\\ {\rm two-loop}: ~~\phantom{\Gamma_{\pi^0\gamma\gamma}}
&=& 8.09 \pm 0.11~{\rm eV}~\
\end{eqnarray}
More precisely, in~\cite{Ananthanarayan:2002kj} the anomalous action was extended to include virtual photons
and the corrections beyond leading order including quark  mass differences and electromagnetic effects were calculated.
In~\cite{Goity:2002nn}, $\pi^0\to 2\gamma$ was analyzed within CHPT combined with the $1/N_C$ expansion,
with the isospin-breaking induced by the mixing of the U(3) states. The NNLO corrections were calculated
in~\cite{Kampf:2009tk} for the two-flavor case and in a modified version of the three-flavor expansion.
All these agree within the uncertainties with the recent experimental results~\cite{PrimEx:2010fvg,PrimEx-II:2020jwd},
\begin{eqnarray}
{\rm PrimEx:}~~\Gamma_{\pi^0\gamma\gamma} &=& 7.82 \pm 0.14 \pm 0.17~{\rm eV}~,\nonumber\\
{\rm PrimEx-II:}~~\Gamma_{\pi^0\gamma\gamma} &=& 7.80 \pm 0.06 \pm 0.11~{\rm eV}~.
\end{eqnarray}
More details on this intriguing decay and its measurements can be found in the review~\cite{Bernstein:2011bx}.

Another interesting prediction of the WZW term relates to  $\gamma\to  3\pi$, namely
\begin{equation}
T(\gamma\to \pi^+\pi^-\pi^0) = -\epsilon_{\mu\nu\alpha\beta} \epsilon^\mu_{} k^\nu_{}
               p_-^\alpha p_+^\beta\, F(s,t,u)~, ~~ {\rm with}~~
F(0,0,0) = F^{3\pi} =  \displaystyle\frac{eN_c}{12\pi^2F_\pi^3} = (9.78\pm 0.05)~{\rm GeV}^{-3}~,
\end{equation}
where the uncertainty stems entirely from the one in $F_\pi$.
This can be tested in the reaction $\gamma \pi\to \pi\pi$ via the so-called Primakoff effect~\cite{Primakoff:1951iae}. \
The first  measurement at Serpuhkov lead to $F^{3\pi} = (12.9 \pm 0.9 \pm 0.5)~{\rm GeV}^{-3}$, in stark 
contrast to the anomaly prediction~\cite{Antipov:1986tp}. It was also shown that one-loop \cite{Bijnens:1989ff}
and two-loop \cite{Hannah:2001ee} corrections to $F^{3\pi}$ are of the order 10\% and 5\%, respectively,
and are thus much too small to explain the discrepancy. A new measurement with the COMPASS experiment,
making use of the improved formalism for  $\gamma \pi\to \pi\pi$ combining dispersion theory with CHPT
constraints \cite{Hoferichter:2012pm},
leads to  $F^{3\pi} = (10.3 \pm 0.6)~{\rm GeV}^{-3}$, perfectly consistent with the chiral anomaly
prediction~\cite{Maltsev}.


\section{CHPT for baryons}\label{sec:baryon}

Consider now the inclusion of matter fields, that is the nucleon doublet ($p,n$) in SU(2) and the
ground state octet ($p,n,\Lambda,\Sigma^0,\Sigma^\pm,\Xi^0,\Xi^-$) in the three-flavor
case, coupled to the Goldstone bosons (and external fields). As we will see, the non-vanishing
baryon mass in the chiral limit induces a scale that needs to be accounted for properly
to achieve a consistent power-counting. All this can only be discussed here in a very rough manner,
and I refer to the reviews for much more details~\cite{Bernard:1995dp,Bernard:2007zu}.

\subsection{Effective Lagrangian}\label{sec:LMB}

First, we restrict ourselves to the two-flavor case and
consider the general structure of the effective pion-nucleon
Lagrangian ${\cal L}_{\pi N}^{\rm eff}$. It contains the pions collected in the
matrix--valued field $U(x)$ and we combine the
proton $(p)$ and the neutron $(n)$ fields in an isospinor $\psi$
\begin{equation}
  \psi = \begin{pmatrix}p\\ n\end{pmatrix}~.
\end{equation}
There is a variety of ways to describe the transformation properties
of the spin-1/2 baryons under chiral SU(2)$_L\times$SU(2)$_R$. All
of them lead to the same physics. However, there is one most convenient
choice, which will be taken here~\cite{Georgi:1984zwz}.
We already know that the pion interactions are of derivative nature
and this needs to be reflected in the effective Lagrangian. This feature can most easily
be incorporated using a non-linear realization of the chiral symmetry. Introducing a matrix-valued
function $K$, the nucleon field transforms as
\begin{equation}
  \psi \to K(L,R,U) \psi~.
\end{equation}
$K$ not only depends on the group
elements $L,R \in {\rm SU(2)}_{L,R}$, but also on the pion field
in a highly non-linear fashion, $K = K(L,R,U)$.
Since $U(x)$ depends on the space-time coordinate $x$, $K$ implicitly
depends on $x$ and therefore the transformations related to $K$
are local. More precisely, $K$ is defined via
$R u = u' K$ with $u^2 (x) = U(x)$ and $U'(x) = R U(x) L^\dagger =
{u'}^2(x)$. The transformation properties of the pion field induce a well-defined
chiral transformation of $u(x)$. This defines $K$
as a non-linear function of $L$, $R$ and $\pi (x)$. $K$ is a realization
of SU(2)$_L\times$SU(2)$_R$,
\begin{equation}
  K = \sqrt{L U^\dagger R^\dagger} R \sqrt U~.
\end{equation}
If one considers infinitesimal transformations, one sees that  the nucleon field is
multiplied with a function of the pion field, or stated differently, chiral
transformations are related to the absorption or emission of pions.
The covariant derivative of the nucleon field is given by
\begin{equation}
D_\mu \psi = \partial_\mu  \psi  + \Gamma_\mu \psi~,~~~
\Gamma_\mu= {1 \over 2} [u^\dagger , \partial_\mu u]
- {i \over 2}  u^\dagger (v_\mu + a_\mu) u
- {i \over 2}  u (v_\mu - a_\mu) u^\dagger~,
\end{equation}
and $D_\mu$ transforms homogeneously under
chiral transformations, $D_\mu' = K D_\mu K^\dagger$.
The object $\Gamma_\mu$ is the so-called chiral connection. It is
a gauge field for the local transformations
$\Gamma_\mu' = K \Gamma_\mu K^\dagger + K \partial_\mu K^\dagger$.
The connection $\Gamma_\mu$ contains one derivative. Therefore,
the lowest order chiral $\pi N$ Lagrangian will start with
terms with $d=1$, very different from the meson case.
One can also form an  object of axial-vector type with one derivative,
\begin{equation}
u_\mu = i (u^\dagger \nabla_\mu u -
u \nabla_\mu u^\dagger) = i \lbrace u^\dagger , \nabla_\mu
u \rbrace = i u^\dagger \nabla_\mu U u^\dagger~,
\end{equation}
which transforms homogeneously, $u_\mu' = K u_\mu K^\dagger $.
The covariant derivative $D_\mu$ and the axial-vector object $u_\mu$
are the basic building blocks for the lowest order effective theory.
Therefore, the LO chiral effective pion-nucleon ($\pi N$) Lagrangian takes the form
\begin{BoxTypeA}[sec4:box1]{}
\begin{equation}\label{eq:LpiN1}
{\cal L}_{\pi N}^{(1)} 
= \bar \psi\, \biggl(i \gamma_\mu D^\mu - \stackrel{\circ}{m} + { \stackrel{\circ}{g_A}  \over 2}
\gamma^\mu \gamma_5 u_\mu\biggr)\, \psi~.
\end{equation}
\end{BoxTypeA}
\noindent
The effective Lagrangian~\eqref{eq:LpiN1} contains two new parameters. These are
the baryon mass $\stackrel{\circ}{m}$ and the axial-vector coupling $\stackrel{\circ}{g_A}$
in the chiral limit, $m= \stackrel{\circ}{m} [1 + {\cal O} ({\cal M}) ]$, 
$g_A = \stackrel{\circ}{g_A} [1 + {\cal O}  ({\cal M}) ]$.
Here, $m = 939$~MeV denotes the physical nucleon mass and $g_A$
the axial-vector coupling measured in neutron $\beta$-decay,
$n \to p e^- \bar \nu_e$, $g_A \simeq 1.27$. The fact that $\stackrel{\circ}{m}$
does not vanish in the chiral limit (or is not small on the typical
scale $\Lambda \simeq 1\,$GeV) will be discussed below.
The occurrence of the constant $\stackrel{\circ}{g_A}$ is all but surprising.
Whereas the vectorial (flavor) SU(2) is protected at zero momentum
transfer~\cite{Ademollo:1964sr}, the axial current is, of course, renormalized.
Together with the LO GB Lagrangian ${\cal L}_{\pi \pi}^{(2)}$ the lowest
order pion-nucleon Lagrangian reads:
\begin{equation}
{\cal L}_1 = {\cal L}_{\pi N}^{(1)} + {\cal L}_{\pi \pi}^{(2)}~.
\end{equation}
To understand the low--energy dimension of ${\cal L}_{\pi N}^{(1)}$,
we have to extend the chiral counting rules  to the
various operators and bilinears involving the nucleon fields,
\begin{eqnarray}\label{eq:powers}
\stackrel{\circ}{m}\!\!\!\! &=& \!\!\!{\cal O}(1)~,~~~ \psi, \, \bar \psi =
{\cal O}(1)~,~~~
D_\mu \psi = {\cal O}(1) \, \, , \, \, \bar \psi \psi  = {\cal O}(1)~,~~~
 \bar \psi \gamma_\mu  \psi
= {\cal O}(1)~,~~~
 \bar \psi \gamma^\mu \gamma_5 \psi
= {\cal O}(1) \, , \nonumber\\
 \bar \psi \sigma^{\mu \nu} \psi\!\!\!\! &=&\!\!\! {\cal O}(1)~,~~~
 \bar \psi \sigma^{\mu \nu} \gamma_5 \psi = {\cal O}(1)~,~~~
  \bar \psi \gamma_5 \psi = {\cal O}(q)~,~~~
 (i D\!\!\!/ - \stackrel{\circ}{m}) \psi = {\cal O}(q)~.
\end{eqnarray}
Here, $q$ denotes a generic nucleon {\em three}--momentum.
Since $\stackrel{\circ}{m}$ is of order one, baryon four-momenta can never be
small on the typical chiral scale. However, the operator
$(i D\!\!\!\!/ - \stackrel{\circ}{m}) \psi$ counts as order ${\cal O}(q)$.
From this LO Lagrangian, one can easily derive the so-called Goldberger-Treiman
relation (GTR)~\cite{Goldberger:1958vp},
\begin{equation}\label{eq:GTR}
g_{\pi NN} = \frac{g_A m}{F_\pi}~,  
\end{equation} 
which is fulfilled in Nature within a few percent.  A quick derivation of the GTR
goes as follows (for simplicity, the various quantities in the chiral limit
are identified with their physical values and we expand in pion fields):
\begin{equation}
  {\cal L}_{\pi N}^{(1)} = \bar\psi \left(i \partial \!\!\!/ - m \right) \psi + \frac{g_A}{2F_\pi}
  \bar\psi \gamma_\mu \gamma_5 \frac{\vec{\tau}}{2} \psi \cdot \partial^\mu \vec{\pi} + \dots~,
\end{equation}  
so that the $N\to N\pi^i$ transition takes the form
\begin{equation}\label{eq:GTR1} 
  T(N\to N\pi^i) = -i \bar{u}(p^\prime) \, \frac{g_A}{2F_\pi} q\!\!\!/ \gamma_5 u(p) \, \tau^i
  = -i \frac{g_A m}{F_\pi} \bar{u}(p^\prime)\,  \gamma_5 u(p)~,  
\end{equation} 
making use of the Dirac equation, $\bar{u}(p^\prime) q\!\!\!/ u(p) = 2m \bar{u}(p^\prime)\gamma_5 u(p)$
and $q = p^\prime-p$.  Now the canonical form of the pion-nucleon coupling is
\begin{equation}\label{eq:GTR2} 
  T(N\to N\pi^i) = -i g_{\pi NN} \,  \bar{u}(p^\prime) \gamma_5 u(p)~.
\end{equation}   
Comparing Eq.~\eqref{eq:GTR2} with Eq.~\eqref{eq:GTR1} leads immediately to the GTR~\eqref{eq:GTR}.
Of course, the GTR is not exact but has corrections, but these are surprisingly small.
Furthermore, from the expansion of the covariant derivative in pion and external fields,
one can read off the so-called Kroll-Ruderman vertex, $(ieg_A/F_\pi) \epsilon\cdot S \, \varepsilon^{a3b}\tau^b$,
with $\epsilon$ the photon polarization vector and $\vec{S} = \vec{\sigma}/2$ the nucleon spin-vector.
This term leads to the low-energy theorems for charged pion photoproduction, see e.g.~\cite{Bernard:1996ti}.

In general, the effective pion-nuclon Lagrangian can be written as,
\begin{equation}
{\cal L}_{\pi N}^{\rm eff} = {\cal L}_{\pi N}^{(1)} + {\cal L}_{\pi N}^{(2)} +  {\cal L}_{\pi N}^{(3)} +  {\cal L}_{\pi N}^{(4)} + \ldots~, 
\end{equation}  
which are all terms required for a complete one-loop analysis in processes invoolving one nucleon in the initial
and the final state. For the construction of the terms beyond $d=1$,
the reader should consult~\cite{Gasser:1987rb,Ecker:1995rk,Fettes:1998ud,Fettes:2000gb}.

In the three-flavor case, we have to deal with the full GB octet~\eqref{eq:U} coupled to the
octet of ground state baryons
\begin{equation}
B =  \begin{pmatrix}
\Sigma^0/\sqrt{2} + \Lambda/\sqrt{6} & \Sigma^+ & p\\
\Sigma^- & -\Sigma^0/\sqrt{2} + \Lambda/\sqrt{6}  & n\\
\Xi^- & \Xi^0 & -2\Lambda / \sqrt{6} \end{pmatrix}~,
\end{equation}
so that the LO meson-baryon Lagrangian takes the form,
\begin{equation}
{\cal L}_{\rm MB}^{(1)} = \langle i \bar{B} \gamma^\mu D_\mu B
- m_0 \bar{B} B + \frac{1}{2} D \bar{B} \gamma^\mu \gamma_5 \lbrace
u_\mu , B \rbrace
+ \frac{1}{2} F \bar{B} \gamma^\mu \gamma_5 [ u_\mu , B ] \rangle~,
\end{equation}
where $m_0$ is for the (average) octet mass in the chiral limit and the
covariant derivative acting on $B$ reads $D_\mu B = \partial_\mu B + [ \Gamma_\mu , B ]$,
$\Gamma_\mu = \frac{1}{2} \bigl\lbrace u^\dagger [ \partial_\mu - i ( v_\mu +
  a_\mu )]u + u [ \partial_\mu - i ( v_\mu - a_\mu )]u^\dagger \bigr\rbrace$. Further,
$D \simeq 0.80$ and $F \simeq 0.46$ are axial-vector coupling constants. Reducing the three-flavor Lagrangiam
to two flavors, one obtains the {\bf matching relation}
\begin{equation}
  g_A = F + D~,
\end{equation}  
which is, of course, exact only at LO. For more such matching relations, see~\cite{Frink:2004ic}.
Note that the LO pion-nucleon (meson-baryon) Lagrangian does not contain any explicit chiral
symmetry breaking terms, these only appear at second order and take the form
\begin{equation}\label{eq:MB2}
{\cal L}_{\rm MB}^{(2)} = b_D\, \langle\bar B \{ \chi_+ , B \}\rangle + b_F \, \langle
\bar B [ \chi_+,B]\rangle + b_0 \, \langle \bar BB\rangle \langle \chi_+ \rangle~,
\end{equation}  
with $\chi_+ = u^\dagger \chi u^\dagger + u \chi^\dagger u$. Note the appearance of three LECs,
from which $b_0$ can not be determined from the baryon masses alone. In fact, evaluating
the baryon masses based on ${\cal L}_{\rm MB}^{(2)}$ (averaged over their isospin components) leads to the famous
Gell-Mann--Okubo relation~\cite{Gell-Mann:1961omu,Okubo:1961jc}, $m_\Sigma + 3 m_\Lambda = 2 (m_N + m_\Xi )$,
which is fulfilled within 0.6$\%$ in Nature.

\subsection{Scales and power counting}\label{sec:baryon PC}

We have already seen in Eq.~\eqref{eq:powers} that the nucleon (baryon) mass is a hard scale of order
one. On the tree level, this poses no problem, and thus LO results can be obtained easily, see
e.g. Eq.~\eqref{eq:WpiN}. However,
matters are different as soon as one considers loops. As stressed in~\cite{Gasser:1987rb}, the appearance
of the nucleon mass in the loop diagrams messes up the power counting, i.e. there is no more one-to-one
correspondence between the expansion in small momenta/masses and loops. This can be understood most
easily by considering the one-dimensional integral
\begin{equation}\label{eq:exloop}
I = \int_{-\infty}^{+\infty} \frac{dq}{2\pi}\frac{q^2}{(l^2+q^2)(h^2+q^2)} = \frac{1}{2(l+h)}~,
\end{equation}
where $l$ and $h$ denote the light and the heavy scale, respectively,  in analogy to  the pion and the 
nucleon mass.  If one assumes that $q$ is of order $l$, then naively this integral scales as ${\cal O}(l)$.
Since we have $l\ll h$, the result can be Taylor-expanded in $l/h$,
\begin{equation}\label{eq:taylor}
\frac{1}{2(l+h)} = \frac{1}{2h}\left(1 - \frac{l}{h} + \frac{l^2}{h^2} + \cdots  \right)~.
\end{equation}
This expression contains both non-analytic (odd powers in $l$) and analytic (even powers in $l$)
terms in $l^2$. Clearly, the first term in the expression~\eqref{eq:taylor} breaks the power counting.
It is also seen that the power-counting breaking terms are analytic in the soft scale, so in principle
they can be absorbed by a proper renormalization prescription. However, such a procedure is
not efficient as it would have to be done in any calculation anew. 
There are essentially three methods to deal with these terms in a more systematic
manner. First, one can perform the Taylor-expansion
on the level of the Lagrangian (or, the action), which automatically leads to a Taylor-expansion
of the propagator containing the hard scale. This is called {\bf Heavy baryon CHPT (HBCHPT)}, first developed
in~\cite{Jenkins:1990jv} and systematically worked out in a path integral formalism in~\cite{Bernard:1992qa}.
Note that in this scheme  the non-analytic pieces are also expanded in one over the baryon mass.
Quite differently, in the other two schemes one starts from the relativistic Lagrangian and
the loop integrals (like the one in~\eqref{eq:exloop}) are directly 
manipulated. In  {\bf Infrared Regularization (IR)} the pertinent loop integrals are rewritten such that
the contributions from the hard scale are separated from the ones from the soft one, and get absorbed
in the renormalization of the LECs of the theory~\cite{Becher:1999he}. Differently to IR, in the 
{\bf Extended-On-Mass-Scheme (EOMS)}, the integrands of the loop integrals are manipulated such that  
only the polynomial terms that break the power counting are subtracted~\cite{Gegelia:1994zz,Fuchs:2003qc}.

Having taken care of the baryon mass, we can extend the argument leading to Eq.~\eqref{eq:amppcmeson}
to the meson-baryon (MB) system using the HBCHPT scheme. The new ingredients are the MB vertices
starting with $d=1$ and the baryon propagator, scaling as $q^{-1}$, so that $\nu = 4L - 2I_M -I_B
+ \sum_d d(N^M_d+N_d^{MB})$. Here, $L$ is the number of loops, $I_M (I_B)$ the number of internal meson (baryon)
lines and $N_d^M, N_d^{MB}$ the number of vertices with dimension $d$ from the meson and meson-baryon Lagrangian,
respectively. We now consider the case with a single baryon line running through the diagram~\cite{Ecker:1994ra},
so that $\sum_d N_d^{MB}= I_B+1$. Utilizing also the topological relation $L = I_B+I_M-\sum_d(N_d^M+N_d^{MB})+1$,
one arrives at
\begin{BoxTypeA}[sec4:box2]{}
\begin{equation}
\nu = 2L+1 +\sum_d (d-2)N_d^M + \sum_d (d-1)N_d^{MB}~.
\end{equation}
\end{BoxTypeA}
\noindent
This has the desired structure as $\nu \geq 2L+1$, so that we have a consistent power counting. The tree level 
has contributions from $\nu=1$ (LO) and $\nu=2$ (NLO), the one-loop corrections are from $\nu=3$ (NNLO)
and $\nu=4$ (N$^3$LO), and so on.

Next, I discuss briefly the three regularization procedures, without going into too much detail. For
definiteness, I will also concentrate on the pion-nucleon system (the two-flavor case). Let us
start with HBCHPT, which by construction is the extreme non-relativistic limit of the relativistic theory and
it shares some similarities with the Foldy-Wouthuysen transformation of the Dirac equation with a heavy field. The nucleon
four-momentum is written as $p_\mu = {m}v_\mu + l_\mu$ in terms of the four-velocity $v_\mu$ with
$v^2=1$ and $v\cdot l \ll {m}$. For ease of notation, I do not differentiate between the mass in the chiral
limit and its physical value, but this needs to be accounted for in actual calculations (and similarly, for
all other quantities). One can now perform a velocity projection of the nucleon spinor $\psi$
with $\psi = \exp(-i {m} v\cdot x)(H+h)$, with $v\!\!\!/ H = H$ and $v\!\!\!/ h = -h$. Using the equation of
motion for the small field $h$, it  can be eliminated completely and the Lagrangian takes the form
\begin{equation}
{\cal L}_{\pi N}^{(1)} = \bar{H} \,\left(i v\cdot D + {g}_A S\cdot u\right) \,H + {\cal O}(1/{m}^2)~,
\end{equation}
in terms of the spin-operator $S_\mu = i\gamma_5\sigma_{\mu\lambda}v^\lambda$. In the rest-frame,
$v_\mu = (1, \vec{0})$ and $\vec{S}=\vec{\sigma}/2$. One sees that the nucleon mass has completely
disappeared from the LO Lagrangian, so that the nucleon propagator takes the form $S(\omega) = -1/(v\cdot l 
+ i\varepsilon)$ with $\varepsilon>0$ and $\omega = v\cdot l$. Also, all Dirac bilinears $\bar\psi \Gamma_\mu \psi$
($\Gamma_\mu = 1, \gamma_\mu, \gamma_5, \ldots$) can be expressed in terms of the four-velocity
$v_\mu$ and the spin-vector $S_\mu$. Further, $v\cdot S=0$ so that the Dirac algebra is very simple in the
extreme non-relatvistic limit. In this approach, the ${\cal O}(q^3)$ contribution to the nucleon
self-energy, that scales as $M_\pi^3$ on dimensional grounds, can be written as
\begin{equation}
\delta m_N^{(3)} = \frac{3g_A^2}{4F_\pi^2} J(0) M_\pi^2~,~~ J(0) = \int \frac{d^Dk}{(2\pi)^D} \frac{1}{(M_\pi^2-k^2-i\varepsilon)
(v\cdot k -i \varepsilon)}~.    
\end{equation}
Evaluating $J(0)$ gives $J(0)=-M_\pi/(8\pi)$, so that we end with the time-honored result $\delta m_N^{(3)} = -(3g_A^2M_\pi^3)/
(32\pi F_\pi^2)$. Thus, in the HBCHPT approach one simply evaluates a standard one-loop integral, here denoted as $J(0)$.
The complete one-loop expansion of the nucleon mass takes the form (after the determination of the LECs)~\cite{Hoferichter:2015hva}
\begin{equation}
m_N =
\underbrace{869.5~{\rm MeV}}_{O(q^0)}\underbrace{+86.5~{\rm MeV}}_{O(q^2)}\underbrace{-15.4~{\rm MeV}}_{O(q^3)}\underbrace{-2.3~{\rm MeV}}_{O(q^4)}
= 938.3~{\rm MeV}~,
\end{equation}
which is a rather well converging series.  The computational simplicity of the extreme non-relativistic limit comes with a price,
namely that in some cases it has an impact on the analytical structure of certain amplitudes, see e.g. Ref.~\cite{Bernard:1996cc}. This, however,
can be remedied by either going to higher orders or by the use of non-expanded kinematical variables. Also, the convergence
of the $1/m$ expansion is found to be slow in some cases (e.g. the neutron electric form factor~\cite{Kubis:2000zd}),
so one usually needs to include the one-loop corrections at NNLO
and not just the leading one-loop terms. As two remarkable findings using HBCHPT we note the abovementioned $1/M_\pi$ singularities
in the nucleon polarizabilities~\cite{Bernard:1991rq} and the modification of the low-energy theorem for neutral pion photoproduction
of the nucleon~\cite{Bernard:1991rt}, see Sect.~\ref{sec:photo}.

In the IR approach, one starts from a manifestly covariant formulation and separates any one-loop integral into an
infrared-singular ($I$) and a regular $(R)$  part. While the latter terms can be expanded and absorbed in the LECs, the
$I$-part has the same analytic structure as the full integral in the low-energy region and is generates the
non-trivial momentum and quark mass dependencies. In fact, summing all heavy baryon diagrams with all internal line insertions yields
the $I$-part of the corresponding relativistic diagram. Consider again the nucleon self-energy, where the corresponding
scalar loop integral takes the form
\begin{equation}
H(p^2) = \frac{1}{i} \int \frac{d^D k}{(2\pi)^D} \frac{1}{[M_\pi^2 -k^2-i\varepsilon]
  [m_N^2 -(p-k)^2-i\varepsilon]}~, ~~ \varepsilon > 0~,
\end{equation}  
which at threshold $p^2 = s_0 = (M_\pi+m_N)^2$ results in $H(s_0) = c(D)\, (M_\pi^{M-3} + m_N^{D-3})(M_\pi + m_N) = I+R$, where 
$c(D)$ is some constant depending on the dimensionality of space-time. As can be seen, the $I$-part is characterized by
fractional powers in the pion mass and is generated by loop momenta of order $M_\pi$, whereas the $R$-part is generated
by internal momenta of the order of the nucleon mass (the large mass scale). These power-counting breaking (PCB) terms are
separated explicitly and since they are analytic in $M_\pi$ can be absorbed in the LECs of the effective Lagrangian.
For the self-energy integral, this splitting can be achieved by rewriting the Feynman parameter integration $\int_0^1 dx ...$
as:
\begin{equation}
H = \biggl\{ \int_0^\infty - \int_1^\infty \biggr\}\, dx
 \int  \frac{d^Dk}{(2\pi)^D} {1 \over [(1-x)(M_\pi^2-k^2-i\varepsilon) + x (m_N^2 -(P-k)^2 -i\varepsilon) ]^2} = I + R~,
\end{equation}  
This integral develops an IR singularity as $M_\pi \to 0$, coming from the low-momentum region of the integration.
Contrary to that, the high-momentum part is free of IR singularities and can thus be expanded in an ordinary power series.
So the desired separation of the soft and hard parts is achieved. This prescription can be easily generalized to arbitrary
one-loop diagrams with $m$ meson and $n$ nucleon propagators~\cite{Becher:1999he}. For the extension to two loops,
see~\cite{Schindler:2007dr}. By construction, the regular part $R$ has a cut along the negative real axis.  While this
is not felt for most processes in the low-energy region~~\cite{Becher:2001hv},  it can lead to distortions in
observables even at low energies that require differentiation with respect to some external
parameter~\cite{Bernard:2002pw} (e.g. the photon virtuality in reactions with one or two external photons).
It is instructive to closer analyze the leading one-loop
correction to the nucleon mass, given in IR by
\begin{equation}
\delta m_N^{(3)}  =  \frac{3g_A^2}{2F_\pi^2} m_N^{} M^2_\pi I(m_N^2)~,~~~ I(m_N^2) = -\mu^2 \biggl(L+\frac{1}{16\pi^2}\log\mu\biggr)  
  +\frac{\mu}{16\pi^2}\biggl\{ \frac{\mu}{2}-\sqrt{4-\mu^2}  
  \arccos\biggl(-\frac{\mu}{2}\biggr) \biggr\} = -\frac{\mu}{16\pi} + O(\mu^2) ~,
\end{equation}  
with $\mu = M_\pi/m_N$ and one recovers the well-known result from the leading order piece of the integral. Note further that
the the loop function $I(m^2_N)$ contains a non-leading divergence which cannot be
absorbed to this order, but will be canceled by an appropriate contact term at fourth order. So one automatically
generates some higher-order terms, that require a further prescription. In the original formulation of Ref.~\cite{Becher:1999he},
the numerators of all one-loop integrals are chirally expanded, whereas in Ref.~\cite{Kubis:2000zd} all terms from the
numerators were kept to accelerate convergence of the chiral expansion.

In the EOMS scheme, which is another covariant approach to baryon CHPT, only the PCB terms
are subtracted, so that the remaining integrals adhere to the power counting discussed above. This can be most easily
understood by considering again the scalar one-loop integral associated to the nucleon self-energy (here for massless
pions):
\begin{eqnarray}\label{eq:Hseries}
H &=& \int \frac{d^Dk}{(2\pi)^Di} \sum_{n=0}^{\infty} \frac{(p^2-m_N^2)^n}{n!}\,
\left[\left(\frac{1}{2p^2}p_\mu\frac{\partial}{\partial p_\mu}\right)^n
\frac{1}{(k^2+i\varepsilon)[k^2-2k\cdot p +(p^2-m_N^2)+i\varepsilon]}\right]_{p^2=m_N^2}\nonumber\\  
&=&  \int \frac{d^Dk}{(2\pi)^Di} \left[\frac{1}{(k^2+i\varepsilon)(k^2 -2k\cdot p+i\varepsilon)}  
  \right]_{p^2=m_N^2}\nonumber\\
&& -\,(p^2-m_N^2)  \left[\frac{1}{2m_N^2}\frac{1}{(k^2 -2k\cdot p+i\varepsilon)^2}
  -\frac{1}{2m_N^2}\frac{1}{(k^2+i\varepsilon)(k^2 -2k\cdot p+i\varepsilon)}
   -\frac{1}{(k^2 + i\varepsilon)(k^2 -2k\cdot p+i\varepsilon)^2}\right]_{p^2=m_N^2} + \ldots\,.
\end{eqnarray}  
The EOMS renormalizations scheme is now defined by the prescription to subtract from the integrand
of $H$ the PCB terms. These terms are always analytic in the small parameters and do
not contain infrared singularities. In the above example we only need to subtract the first
term. All the higher-order terms contain infrared singularities. For example, the last term
of the second coefficient in  Eq.~\eqref{eq:Hseries} would generate a behavior $\sim k^3/k^4$ of the
integrand for $D=4$. Thus in the EOMS approach one has $H^r = H-H^{\rm sub}$
with:
\begin{equation}
H^{\rm sub} = -i \int \frac{d^Dk}{(2\pi)^D} \left[\frac{1}{(k^2+i\varepsilon)(k^2 -2k\cdot p+i\varepsilon)}
  \right]_{p^2=m_N^2}~,
\end{equation}
which is indeed the first term in the expansion. These PCB terms can be absorbed in the LECs of the chiral
$\pi N$ Lagrangian. For the case with non-vanishing pion masses, see~\cite{Fuchs:2003qc}. The EOMS
approach looks very appealing and has become popular lately, but it is not free of deficencies either.
While in HBCHPT and the IR approach the chiral Ward identities are fulfilled by construction, special
care has to be taken in EOMS not to break chiral symmetry. For a detailed discussion, the reader is
referred to~\cite{Meissner:2022cbi}. Note also that from the EOMS scheme, one can recover the HBCHPT
results after subtracting all PCB terms and expanding in $1/m_N$. The IR approach can also be obtained from
EOMS, see~\cite{Schindler:2003xv}.

\subsection{Assorted applications}\label{sec:baryonapp}

Similar to the GB section~\ref{sec:meson}, I only discuss a few assorted applications here, more precisely, two cases
were the chiral loops indeed led to modifications of time-honored LETs and thus contributed to a better understanding
of the chiral pion-nucleon dynamics.

\subsubsection{Neutral pion photoproduction off the nucleon}
\label{sec:photo}

\begin{figure}[htb!]
	\centering
        \includegraphics[width=.45\textwidth]{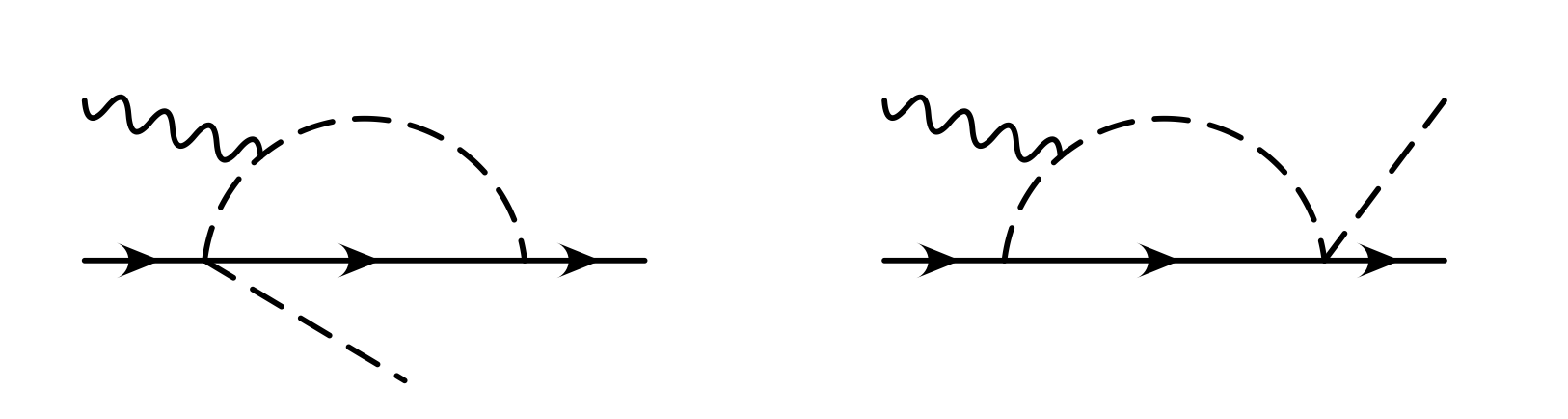}
        \caption{Left: The triangle diagram in neutral pion photoproduction off nucleons. Solid, dashed and wiggly lines refer
        to  nucleons, pions and photons, in order.}
	\label{fig:E0+}
\end{figure}

We consider the process $\gamma (k) +N (p_1) \to \pi^0 (k)+ N(p_2) ~ (N=p,n)$
at threshold, i.e., for vanishing three-momentum of the pion in the nucleon rest-frame, $\vec{k}=0$.
In the laboratory frame, this corresponds to a photon energy $E_\gamma = 144.7\,$MeV.
At threshold, only the electric dipole amplitude $E_{0+}$ survives  (here, the subscripts
refer to the pion angular momentum $\ell=0$, i.e. the S-wave, and $+$ denotes 
the total angular momentum, that is $j=\ell+1/2$) so that the differential cross section is given by
\begin{equation}
\frac{d\sigma}{d\Omega} = \frac{|\vec{q}\,|}{|\vec{k}\,|}\,|E_{0+}|^2~.
\end{equation}
Note that $E_{0+}$ is complex-valued, but it is real at threshold and the imaginary part
is very tiny below the opeing of the $\pi^+ n$ channel at $E_\gamma=151.4\,$MeV.
The only quantity contributing to $E_{0+}$ with non-zero chiral dimension is $M_\pi$. In the usual conventions,
$E_{0+}$ has physical dimension $-1$ and it can therefore be written as
\begin{equation}
E_{0+} = {e g_A \over F}~A\left( {M_\pi\over m_N},{M_\pi\over F}\right)~,
\end{equation}
where $F$ is the pion decay constant in the chiral limit and $m_N=(m_p+m_n)/2$ the nucleon mass.
The dimensionless amplitude $A$ can now be expressed as a power series in $M_\pi$.  
Before CHPT the LETs were derived based on current algebra and had the form (expressing all
quantities in terms of their physical values)~\cite{DeBaenst:1970dqx,Vainshtein:1972ih}
\begin{eqnarray}\label{eq:wrong}
E_{0+}(\pi^0 p) &=&  -\frac{e g_A}{8\pi F_\pi}\left[ \frac{M_\pi}{m_N} - \frac{3+\kappa_p}{2}\frac{M_\pi^2}{m_N^2}
                +{\cal O}(M_\pi^3) \right]~, \nonumber\\
E_{0+}(\pi^0 n) &=&  -\frac{e g_A}{8\pi F_\pi}\left[ \frac{3+\kappa_n}{2}\frac{M_\pi^2}{m_N^2} +{\cal O}(M_\pi^3) \right]~, 
\end{eqnarray}
with  $\kappa_n = -1.913$  anomalous magnetic moment of the neutron.
These expressions  are  intriguing as not only the leading term $\sim M_\pi$ could be derived (which vanishes in case of the
neutron since $Z_n=0$) but also the subleading term $\sim M_\pi^2$. However, the first one-loop analysis within baryon CHPT
revealed that there are corrections at order $M_\pi^2$ that originate from the so-called triangle diagram, see Fig.~\ref{fig:E0+},
so that the correct form of the LETs reads~\cite{Bernard:1991rt}
\begin{eqnarray}
E_{0+}(\pi^0 p) &=&  -\frac{e g_A}{8\pi F_\pi}\left[ \frac{M_\pi}{m_N}  - \left(\frac{3+\kappa_p}{2}+\frac{1}{16\pi^2}\right)
 \frac{M_\pi^2}{m_N^2} +{\cal O}(M_\pi^3)\right]~, \nonumber\\
E_{0+}(\pi^0 n) &=&  -\frac{e g_A}{8\pi F_\pi}\left[ \left(\frac{3+\kappa_n}{2}+\frac{1}{16F_\pi^2}\right)
 \frac{M_\pi^2}{m_N^2} +{\cal O}(M_\pi^3)
\right]~.
\end{eqnarray}
This new term at second order in the pion mass is generated from the non-analyticity of the triangle
loop diagram when $M_\pi$ goes to zero,  as it has an IR singularity in this limit. This is yet another manifestation of the 
intricate physics in the chiral  limit. As this correction is numerically large,
the contributions $\sim M_\pi^3$ have to be considered.  These have been worked out in Ref.~\cite{Bernard:1994gm}
and in a nice interplay between experiment and
theory, the S-wave amplitude $E_{0+}(\pi^0 p)$ could be pinned down  quite accurately in the threshold region, exhibiting a
cusp at the opening of the $\pi^+ n$ threshold just $6.8$~MeV above the $\pi^0 p$ threshold.  Combining now
CHPT with dispersion relations to analyze the most precise threshold data from the Mainz Microtron 
MAMI-B~\cite{Schmidt:2001vg} led to $E_{0+} = -1.19 \cdot 10^{-3}/M_\pi$, which is about a factor of two smaller
than the prediction from the incorrect LEC~\eqref{eq:wrong}. Another intriguing CHPT prediction is that the
neutron S-wave multipole $E_{0+}(\pi^0 n)$ at threshold is about twice as large in magnitude as the one
for the proton~\cite{Bernard:1994gm},
which completely defies the intuition based on the classical dipole picture, see e.g. Ref.~\cite{Ericson:1988gk}.
This prediction has not been tested accurately, it is, however, consistent with the existing data for neutral
pion photoproduction off the deuteron~\cite{Beane:1997iv}. It has also been shown that the reaction 
$\gamma + {}^3{\rm He}\to \pi^0 + {}^3{\rm He}$ would be good testing ground for this intriguing
prediction~\cite{Lenkewitz:2012ixw}. Further developments with respect to neutral pion productions of nucleons
(inclusion of higher partial waves, isospin breaking, calculations within the EOMS scheme, the range of
applicability of the chiral expansion, P-wave LETs, etc.) are reviewed in the talk~\cite{Meissner:2022yyi}.

\subsubsection{The nucleon axial radius from pion electroproduction}

The matrix-element of the SU(2) isovector axial quark current
$A_\mu^a = \bar{q} \, \gamma_\mu \, \gamma_5 \, (\tau^a/2) \, q$
between nucleon states is given by 
\begin{equation}
\label{current}
\langle N(p') \, | A_\mu^a \, | N(p)\rangle = \bar{u}(p') \,
\left[ \gamma_\mu \, G_A (k^2) + {(p' -p)_\mu \over 2m_N}\,
G_P (k^2) \, \right] \, \gamma_5 {\tau^a \over 2} \, u(p)~,
\end{equation}
with $k^2=(p'-p)^2\leq 0$ the invariant momentum transfer squared and
$m_N = (m_p + m_n)/2$ the nucleon mass. Further, $G_A(k^2)$ is the axial
form factor and  $G_P (k^2)$ the induced pseudoscalar form factor.
Here, we will only consider  $G_A(k^2)$.
It can be Taylor-expanded for small photon
virtualities  as (see e.g. the review~\cite{Bernard:2001rs})
\begin{equation}
\label{eq:taylorGA}
G_A (t) = g_A \, \left( 1 + \frac{1}{6} \langle r_A^2 \rangle \, k^2 + {\cal O}(k^4) \right)~,
\end{equation}
with $g_A$ the axial-vector coupling constant, and $\langle r_A^2 \rangle^{1/2} \simeq 0.67\,$fm the nucleon axial radius. 
This is yet another scale that characterizes the nucleon at low energies, which differs sizeably from the electromagnetic
size, as e.g. the charge radius of the proton is $r_E^p =0.84\,$fm. The axial form factor (or the axial radius) can be
determined from (anti)neutrino-nucleon scattering or
pion electroproduction off the nucleon. Before CHPT tools became available, the so determined axial radius
differed by about 10\%, $\langle r_A^2 \rangle_{\nu-{\rm scatt.}}^{1/2} = (0.666 \pm 0.014)\,$fm
versus $\langle r_A^2 \rangle_{{\rm elprod.}}^{1/2} = (0.639 \pm 0.010)\,$fm, see Ref.~\cite{Bernard:2001rs}.
The analysis of the electroproduction data was based on  the LET of Nambu, Luri\'e
and Shrauner~\cite{Nambu:1962ilv,Nambu:1962lbq} for the electric dipole amplitude (for massless pions),
\begin{equation}\label{eq:LET-NLS}
E_{0+}^{(-)} (M_\pi =0, k^2) = 
{e g_A \over 8\pi F_\pi}
\, \biggl\{ 1 + {k^2 \over 6} \langle r_A^2 \rangle + {k^2 \over 4m_N}
\biggl(\kappa_v + \frac{1}{2}\biggr) + {\cal O}(k^3) \biggr\}~,
\end{equation}
with $\kappa_v$ the nucleon isovector anomalous magnetic moment, the superscript (-) refers
to the isospin amplitude $\sim [\tau_a,\tau_3]$ and $a$ is the pion isospin index. This had to be supplemented with some
prescription to include the finite pion mass. For a detailed discussion of pion electroproduction
in the framework of HBCHPT, see Ref.~\cite{Bernard:1993bq}.
In Ref.~\cite{Bernard:1992ys}, one-loop corrections to the amplitude $E_{0+}^{(-)}$ were
analyzed and it was found that the coefficient of the $k^2$-term in Eq.~\eqref{eq:LET-NLS}  receives an 
extra contribution,
\begin{equation}
 1 + {k^2 \over 6} \langle r_A^2 \rangle + {k^2 \over 4m_N}
\biggl(\kappa_v + \frac{1}{2}\biggr) + \frac{k^2}{128F_\pi^2}
\biggl( 1 - \frac{12}{\pi^2} \biggr) ~.
\end{equation}
The last term in this equation constitutes a model-independent
contribution at order $k^2$ not modified by higher loop or 
contact terms, which are suppressed by powers of $M_\pi$.
Formally, it appears because one cannot interchange the limits
$M_\pi \to 0$ (chiral limit) and $k^2 \to 0$ (photon point).
Therefore, what was believed to be the axial radius determined 
in pion electroproduction was nothing but the modified radius
\begin{equation}
\langle \tilde{r}_A^2 \rangle = \langle {r}_A^2 \rangle +
\frac{3}{64 F_\pi^2} \biggl( 1 - \frac{12}{\pi^2} \biggr) ~.
\end{equation}
The numerical value of this additional term is  $\Delta \langle {r}_A^2 \rangle = 
\langle \tilde{r}_A^2 \rangle - \langle {r}_A^2 \rangle
= -0.0456\,{\rm fm}^2$, which closes the abovementioned gap between the determinations
from neutrino scattering and pion electroproduction. The ${\cal O}(q^4)$ corrections
to the radius shift have also been
calculated~\cite{Bernard:1995dp} and only modify the leading ${\cal O}(q^3)$ result by about 10\%.
For a recent determination of the axial form factor from antineutrino-proton scattering, see Ref.~\cite{MINERvA:2023avz},
and a detailed discussion of extracting this obervable from lattice QCD data constrained by chiral
symmetry is given in Ref.~\cite{Lutz:2020dfi}.

\section{Conclusions}
\label{sec:conclusions}
Chiral perturbation theory is the effective field theory of QCD for the light
quark sector and below the chiral symmetry breaking scale $\Lambda_\chi \simeq 1\,$GeV.
Its basic degrees of freedom are the pions, the kaons and the eta meson that can be identified
with the Goldstone bosons of the spontaneously broken chiral symmetry of QCD.
This chiral symmetry is further broken explicitly through the small quark masses
$m_u,m_d,m_s$ giving mass to the Goldstone bosons. Further, it is broken
anomalously through renormalization, which is encoded in the so-called Wess-Zumino-Witten
term. Within CHPT, any S-matrix element
or transition current can be expanded in small meson momenta and Goldstone boson masses,
collected in a small parameter $q$. In the meson sector, only operators with even powers
of $q$ are allowed (Lorentz invariance, parity), so that there is a one-to-one correspondence
between the expansion in powers of $q^2$ and the loop expansion. This comes with the
appearance of low-energy constants, that on the one hand parameterize the short-distance
physics and on the other hand absorb the UV divergences of the loop diagrams
via order-by-order renormalization. Applications of this method discussed here are pion-pion scattering
as the purest process in chiral dynamics,
quark mass ratios from Goldstone boson mass ratios and tests of the chiral anomaly encoded in the
Wess-Zumino-Witten term. Matter fields like the
nucleon in the two-flavor case (or the baryon ground state octet for three flavors)
can also be included in the effective Lagrangian, that then also contains odd
powers of $q$ due to the fermionic nature of the baryons. The main complication when including
baryons is the large scale introduced by the baryon mass, $m_B \simeq \Lambda_\chi$. This
messes up the power counting and thus requires a special treatment to recover a
consistent power counting. This can  either be achieved by considering the
baryons as extremely heavy or by  manipulating the loop integrals in a way that allows to
cleanly separate the power-counting breaking terms and absorb them in low-energy constants.
As particular applications neutral pion photoproduction off nucleons and the nucleon
axial radius in pion electroproduction were discussed, leading to new insights of the
chiral dynamics of QCD.


\begin{ack}[Acknowledgments]%
I thank all my collaborators who have shaped my understanding of the topics discussed here,
especially V\'eronique Bernard, J\"urg Gasser and Norbert Kaiser. I am very grateful to
 J\"urg Gasser and Feng-Kun Guo for a careful reading of the manuscript.
This work was supported in part by the Deutsche Forschungsgemeinschaft (DFG, German Research
Foundation) and the NSFC through the funds provided to the Sino-German Collaborative  
Research Center TRR~110 ``Symmetries and the Emergence of Structure in QCD''
(DFG Project-ID 196253076 - TRR 110, NSFC Grant No. 12070131001), by the Chinese Academy
of Sciences  through the President's International Fellowship Initiative (PIFI) under Grant No. 2025PD0022,
by the European Research Council (ERC) under the European Union's
Horizon 2020 research and innovation programme (AdG EXOTIC, grant agreement No. 101018170)
and  by the MKW NRW under the funding code NW21-024-A. 
\end{ack}

\seealso{Effective field theory; Chiral symmetry (breaking); Lattice field theory; Nucleon mass: trace anomaly and sigma
  terms; Pions decays; Kaon physics; Strong CP problem, Theta term and QCD topological properties}

\bibliographystyle{Numbered-Style} 
\bibliography{refugm}

\end{document}